\documentclass [12pt]{article}
\usepackage{geometry} 
\usepackage[cp1251]{inputenc}
\usepackage[english,russian,ukrainian]{babel}
\usepackage{graphicx}

\begin{document}

\begin{center}
\textbf{Analytical forms of the wave function and the asymmetry
for polarization characteristics of the deuteron}
\end{center}

\begin{center}
V.I. Zhaba
\end{center}

\begin{center}
\textit{Uzhgorod National University, Department of Theoretical
Physics,}
\end{center}

\begin{center}
\textit{54, Voloshyna St., Uzhgorod, UA-88000, Ukraine}
\end{center}

\begin{center}
\textit{e-mail: viktorzh@meta.ua}
\end{center}

The basic analytical forms for the deuteron wave function (DWF) in
coordinate representation have been reviewed. The asymptotic
behaviors of DWF near the origin of coordinates have been
analyzed. New analytical forms DWF as a product of exponential
function $r^{n}$ by the sum of the exponential terms
$A_{i}$\textit{*exp(-a}$_{i}*r^{3})$ are applied for calculation
polarization characteristics of the deuteron. Numerical
calculations have been done for realistic phenomenological
potential Nijmegen groups - Reid93. In paper the results of
angular asymmetry for deuteron vector$ t_{10}$, $t_{11}$ and
tensor$ t_{20}$, $t_{21}$, $t_{22}$ polarizations are submitted.
Except for angular asymmetry, is present as well momentum
asymmetry for deuteron vector $t_{1i}$ polarizations. Influence of
approximation DWF in coordinate representation on the subsequent
results of calculations tensor polarization $t_{20}$ is
investigated. Comparison of values $t_{20}$ is made at
\textit{$\theta $}=70 degree if to apply four different
approximations DWF for same NN potential Reid93. At comparison of
the received theoretical values $t_{20}$ with experimental data of
world collaborations and reviews the good coordination for area of
values momentum $p$=1-4 fm$^{-1}$ is observed. Within the
framework of a method of invariant amplitude have been calculated
spin observables in backward elastic \textit{dp}- scattering -
tensor analyzing power $T_{20}$ and polarization transfer \textit{
$\kappa $}$_{0}$. Are compared values for \textit{$\kappa $}$_{0}$
and correlation \textit{$\kappa $}$_{0}-T_{20}$ with experimental
data's. In a wide range of momentums $p$ and angles
\textit{$\theta $} scattering have been presented asymmetry for
tensor analyzing power $T_{20}$ and $T_{22}$, which characterize
photoproduction for negative $\pi $- meson from deuteron in
reaction \textit{$\gamma $(d},$\pi ^{-})$\textit{pp}. Symmetry of
values $T_{20}$ and $T_{22}$ concerning a angle 90 degree is
observed. Ratio $R$ for vector $P_{x}$ and tensor $P_{xz}$
polarizations is characterized by angular asymmetry. So, near to
``$A_{y}$ puzzle'' remain actual theoretical and experimental
researches of other polarization characteristics for processes at
participation deuteron for which is present both angular, and
momentum asymmetry. The obtained results of the deuteron vector
and tensor polarizations $t_{ij}(p)$ give some information about
the electromagnetic structure of the deuteron and the differential
cross section of double scattering.


\begin{center}
\textbf{Аналітичні форми хвильової функції і асиметрія поляризаційних характеристик дейтрона}
\end{center}

\begin{center}
В.І. Жаба
\end{center}

\begin{center}
\textit{Ужгородський національний університет, кафедра теоретичної
фізики,}
\end{center}

\begin{center}
\textit{вул. Волошина, 54, Ужгород, 88000, Україна}
\end{center}

Проведено огляд та аналіз асимптотичної поведінки основних
аналітичних форм хвильової функції дейтрона (ХФД) в координатному
представленні. Для розрахунку поляризаційних характеристик
дейтрона застосовано нові аналітичні форми ХФД у вигляді добутку
степеневої функції $r^{n}$ на суму експоненціальних членів
$A_{i}$\textit{$\cdot $exp(-a}$_{i}$\textit{$\cdot $r}$^{3})$. Для
чисельних обчислень використано реалістичний феноменологічний
потенціал Reid93. У роботі представлені результати кутової
асиметрії для векторних $t_{10}$, $t_{11}$ і тензорних $t_{20}$,
$t_{21}$, $t_{22}$ дейтронних поляризацій. Окрім кутової
асиметрії, наявна й імпульсна асиметрія для векторних $t_{1i}$
дейтронних поляризацій. Досліджено вплив чотирьох апроксимація ХФД
на результати чисельних розрахунків тензорної поляризації
$t_{20}$. У рамках методу інваріантної амплітуди проведено
розрахунок тензорної аналізуючої здатності $T_{20}$ і
поляризаційної передачі \textit{$\kappa $}$_{0}$, які є спіновими
спостережуваними в пружному \textit{dp}- розсіянні назад.
Порівняються значення $t_{20}$, \textit{$\kappa $}$_{0}$ та
кореляції \textit{$\kappa $}$_{0}-T_{20}$ з експериментальними
даними провідних колаборацій та оглядів. Відносно кута 90$^{0}$
cпостерігається симетрія аналізуючих здатностей $T_{20}$ і
$T_{22}$, які характеризують фотонародження негативного $\pi $-
мезона в реакції \textit{$\gamma $(d},$\pi ^{-})$\textit{pp}.
Відношення $R$ для векторної $P_{x}$ і тензорної $P_{xz}$
поляризацій характеризується кутовою асиметрією. Поряд з ``$A_{y}$
загадкою'' залишаються актуальними теоретичні та експериментальні
дослідження групи поляризаційних характеристик для процесів за
участю дейтрона, для яких наявна як кутова, так й імпульсна
асиметрія. Отримані результати для векторної і тензорної
дейтронних поляризацій $t_{ij}(p)$ дають певну інформацію про
електромагнітну структуру дейтрона і диференціальний переріз
подвійного розсіяння.

\textbf{Ключові слова}: дейтрон, хвильова функція, аналітична
форма, поляризація, асиметрія.

PACS number(s): 13.40.Gp, 13.88.+e, 21.45.Bc, 03.65.Nk

\textbf{Вступ}

Дейтрон є найпростішім ядром. Він складається з двох сильно
взаємодіючих елементарних частинок - протона і нейтрона. Простота
і наочність будови дейтрона робить його зручною лабораторією для
вивчення і моделювання нуклон-нуклонних сил. На сьогодні дейтрон
добре вивчений експериментально і теоретично.

Експериментально визначені значення статичних характеристик
дейтрона добре узгоджуються з експериментальними даними. Однак
незважаючи на це, існують певні теоретичні неузгодженості і
проблеми. Так у деяких роботах одна (наприклад, для OBEP
\cite{Machleidt1987}, Bonn \cite{Machleidt2001} потенціалів) або
обидві (для Soft core Reid68 \cite{Reid1968}, Moscow
\cite{Kukulin1998}, ренормалізованих кіральних ОРЕ та ТРЕ
\cite{Arriola2007} потенціалів) компоненти радіальної хвильової
функції дейтрона (ХФД) в координатному представленні мають вузли
поблизу початку координат. Існування вузлів у хвильових функціях
основного і єдиного стану дейтрона свідчить про неузгодженості і
неточності в реалізації чисельних алгоритмів в розв'язанні
подібних задач. Це також пов'язано з особливостями потенціальних
моделей для опису дейтрона. Вплив вибору чисельних алгоритмів на
розв'язки задач приведено в роботах \cite{Haysak2009, Haysak2014,
Bokhinyuk2012}. Крім того, слід зазначити що хвильова функція
дейтрона в імпульсному представленні у науковій літературі
представлена неоднозначно. Зокрема, в S- компоненти
\cite{Garcon2001, Veerasamy2011, Fukukawa2015} (або в S- і D-
компонент \cite{Loiseau1987, Gross2010}) присутній надлишковий
вузол в середині інтервалу значень для імпульсу.

Крім надлишкових вузлів хвильової функції дейтрона, до
нерозв'язаних проблем нуклон-нуклонних взаємодій також відносять
так звану ``$A_{y}$ puzzle (загадка)'' \cite{Glockle1996,
Huber1998} - це енергетична поведінка поляризаційного параметру
асиметрії $A_{y}$. У роботі \cite{Huber1998} систематично
досліджено які саме покращення в описі величини $A_{y}$ можуть
бути отримані шляхом змін NN потенціалу.

Також слід відмітити, що такі потенціали нуклон-нуклонної
взаємодії, як Боннський, Московський, потенціали Неймегенської
групи (NijmI, NijmII, Nijm93, Reid93 \cite{stoks1994, Swart1995}),
Argonne v18 \cite{Wiringa1995}, Парижський, NLO, NNLO and
N$^{3}$LO, Idaho N$^{3}$LO чи Oxford мають досить непросту
структуру і громіздкий запис. Параметри потенціальних моделей
оптимізовані таким чином, що мінімізовано значення \textit{$\chi
$}$^{2}$ у прямій підгонці до даних. Для потенціалу Nijm92pp
величина \textit{$\chi $}$^{2}/N_{рр}$ становила 1.4. Наступне
удосконалення потенціалу Nijm78 для \textit{np} даних дало модель
Nijm93:\textit{ $\chi $}$^{2}/N_{pp}$=1.8 для 1787 \textit{pp} і
\textit{$\chi $}$^{2}/N_{np}$=1.9 для 2514 \textit{np} даних,
тобто \textit{$\chi $}$^{2}/N_{data}$=1.87. Для потенціалів Nijm I
і NijmII величина \textit{$\chi $}$^{2}/N_{data}$=1.03.
Оригінальний потенціал Рейда Reid68 \cite{Reid1968} був
параметризований Неймегенською групою на основі фазового аналізу і
отримав назву як оновлена регуляризована версія - Reid93.
Параметризація була проведена для 50 параметрів потенціалу,
причому значення \textit{$\chi $}$^{2}/N_{data}$=1.03
\cite{stoks1994, Swart1995}. Потенціал Argonne v18
\cite{Wiringa1995} з 40 регульованими параметрами дає величину
\textit{$\chi $}$^{2}/N_{data}$=1.09 для 4301 \textit{pp} і
\textit{np} даних в області енергій 0-350~MeВ. Для потенціалу
CD-Bonn \cite{Machleidt2001} величина \textit{$\chi
$}$^{2}/N_{data}$ становить 1.01 для 2932 \textit{pp} даних і 1.02
для 3058 \textit{np} даних.

До сучасних феноменологічних нуклон-нуклонних потенціалів
взаємодії відносяться наступні: залежний від заряду потенціал
CD-Bonn \cite{Machleidt2001}, потенціал Moscow \cite{Kukulin2001},
кіральний потенціал Айдахо \cite{Entem2002}, релятивістська
оптична модель \cite{Knyr2006} на базі Московського потенціалу,
бразильський релятивістський потенціал двопіонного обміну
$O(q^{4})$ \cite{Rocha2007}, локальний нуклон-нуклонний потенціал,
розширений з точки зору ортогональних проекторів
\cite{Kamuntavicius2010}, потенціали Неймегенської групи
\cite{stoks1994, Swart1995} та ін.

Крім цього, ХФД в координатному представленні може бути
представлена таблично: через відповідні масиви значень радіальних
хвильових функцій. Іноді при чисельних розрахунках оперувати
такими масивами чисел доволі складно і незручно. І текст
програмного коду для чисельних розрахунків є громіздкий,
перевантажений і нечитабельним. Тому є доцільним отримання більш
простих аналітичних форм представлення ХФД. В подальшому по них
можна розрахувати формфактори і тензорну поляризацію, що
характеризують структуру дейтрона. ХФД у зручній формі необхідні
для використанні у розрахунках поляризаційних характеристик
дейтрона, а також для оцінки теоретичних значень спінових
спостережуваних в \textit{dp}- розсіянні \cite{Ladygin1997}.

Основними завданнями дослідження даної роботи є огляд основних
аналітичних форм ХФД в координатному представленні та розрахунок
поляризаційних характеристик та аналіз їх асиметрії в процесах за
участю дейтрона.

\textbf{1. Аналітичні форми ХФД}

Хвильова функція описує квантову-механічну систему і є основною
характеристикою мікрооб'єктів. Знання ХФД дозволяє одержувати
максимальну інформацію про зв'язану систему нейтрон-протон і
теоретично обчислювати та передбачати характеристики, що в свою
чергу досліджуються на експерименті. ХФД записується у виді суми
хвильових функцій $^{3}$S$_{1}$- і $^{3}$D$_{1}$- станів
\cite{Blatt1958}

\[
\Psi _d = \psi _S + \psi _D = \frac{u(r)}{r}Y_{101}^1 +
\frac{w(r)}{r}Y_{121}^1 ,
\]

де $u(r)$ і $w(r)$ - радіальні ХФД для S- і D- станів для
орбітальних моментів $l$=0 i $l$=2; $Y_{JLS}^M (\theta ,\phi )$ -
сферичні гармоніки, які визначаються орбітальним моментом $L$,
спіном $S$, повним моментом кількості руху $J=L+S$ та його
проекцією $M$. Для дейтрона повний спін $S$=1, а сумарний момент
$J=M=S$=1. Радіальні ХФД $u(r)$ і $w(r)$ отримують як розв'язки
системи зв'язаних рівнянь Шредінгера \cite{Haysak2009}

\[
\left\{ {\begin{array}{l}
 \frac{d^2u(r)}{dr^2} + \left( { - k^2 - U_1 (r)} \right)u(r) = \sqrt 8 U_3
(r)w(r), \\
 \frac{d^2w(r)}{dr^2} + \left( { - k^2 - \frac{6}{r^2} - U_2 (r)}
\right)w(r) = \sqrt 8 U_3 (r)u(r), \\
 \end{array}} \right.
\]

де $U_i (r) = \frac{2\mu }{\hbar ^2}V_i (r)$ - віднормовані
потенціали; $U_1 $, $U_2 $ - потенціали каналів $l$=0;2; $U_3 $ -
тензорна компонента нуклон-нуклонної взаємодії; $k^2 = -
\frac{2mE}{\hbar ^2}$ - хвильовий вектор; \textit{$\mu $} -
приведена маса. Зв'язуючим потенціалом такої системи є тензорна
частина NN- потенціалу взаємодії.

При описі ХФД в координатному представленні використовують такі
терміни, як ``аналітична форма'', її ``апроксимація'' або
``параметризація''. В першу чергу термін ``аналітична форма''
використовується в якості отриманого розв'язку системи зв'язаних
рівнянь Шредінгера. Пізніше в роботах цей вираз використовується
для позначення записів, отриманих в результаті наближення ХФД. У
детальному огляді \cite{Zhaba0} аналітичні форми ХФД представлені
відповідно до позначень, зазначених у цитованій літературі. Слід
звернути увагу на те, що найбільш вживаними і використовуваними
аналітичними формами є наступні.

1) Аналітична форма Гюлтен-Сугавара типу була запропонована у виді
\cite{donnachie1962}

\begin{equation}
\label{eq1} \left\{ {\begin{array}{l}
 u_g (r) = \cos \varepsilon _g \left[ {1 - e^{ - \beta (x - x_C )}}
\right]e^{ - x}; \\
 w_g (r) = \sin \varepsilon _g \left[ {1 - e^{ - \gamma (x - x_C )}}
\right]^2e^{ - x}\left[ {1 + \frac{3(1 - e^{ - \gamma x})}{x} +
\frac{3(1 -
e^{ - \gamma x})^2}{x^2}} \right]; \\
 \end{array}} \right.
\end{equation}

де $N^{2}$=7.6579$\times $10$^{-12}$cm$^{-1}$; $x = \alpha r$;
$x_C = \alpha r_C $; $\alpha $=0.2316 fm$^{-1}$; $r_{C}$ - радіус
твердої серцевини. Два набори значень вибирають для імовірності D-
стану

$\beta$=7.961; $\gamma$= 3.798; sin$\varepsilon _{g}$=0.02666 для
4{\%} D- стану;

$\beta$=7.451; $\gamma$ = 4.799; sin$\varepsilon _{g}$=0.02486 для
6{\%} D- стану.

2) У роботі \cite{bialkowski1964} передбачено, що дійсна хвильова
функція - це сума ``зовнішньої'' частини, знайденої для відомого
ОРЕ потенціалу, і ``внутрішньої'' частини. ``Зовнішня'' частина
повільніше, ніж ``внутрішня'' частина зникає по експоненті між
однією і двома масами піона. Сам набір ХФД задається парами
доданків:

\begin{equation}
\label{eq2} \left\{ {\begin{array}{l}
 u(r) = u_{outer} + u_{inner} , \\
 w(r) = w_{outer} + w_{inner} , \\
 \end{array}} \right.
\end{equation}

\noindent де ``зовнішня'' та ``внутрішня'' частини (``outer'' та
``inner'' відповідно) рівні

\[
\begin{array}{l}
 u_{outer} = Ae^{ - \kappa r}\left[ {1 + \int {\frac{\rho ^ +
(\alpha )e^{ - \alpha r}dr}{\alpha (\alpha + 2\kappa )} + H\int
{\frac{\rho ^ - (\alpha )e^{ - \alpha r}dr}{\alpha (\alpha +
2\kappa )}} } }
\right], \\
 w_{outer} = Ae^{ - \kappa r}\left[ {H + \int {\frac{\sigma ^ +
(\alpha )e^{ - \alpha r}dr}{\alpha (\alpha + 2\kappa )} + H\int
{\frac{\sigma ^ - (\alpha )e^{ - \alpha r}dr}{\alpha (\alpha +
2\kappa )}} }
} \right], \\
 \end{array}
\]

\[
\left\{ {\begin{array}{l}
 u_{inner} = Ae^{ - \kappa r}\left[ {\gamma _1 e^{ - \xi _1 r} + \gamma _2
e^{ - \xi _2 r}} \right], \\
 w_{inner} = Ae^{ - \kappa r}\left[ {\gamma _3 e^{ - \xi _1 r} + \gamma _4
e^{ - \xi _2 r}} \right]. \\
 \end{array}} \right.
\]

3) Сепарабельний тензорний потенціал Yamaguchi генерує ХФД в
імпульсному представленні. Перетворення Фур'є трансформує таку
хвильову функцію в координатне представлення \cite{Burnap1970}

\begin{equation}
\label{eq3} \left\{ {\begin{array}{l}
 u(r) = e^{ - \alpha r} - e^{ - \beta r}, \\
 w(r) = \eta \left[ {\left( {1 + \frac{3}{\alpha r} + \frac{3}{\alpha
^2r^2}} \right)e^{ - \alpha r} + } \right. \\
 + \left. {\left( {\frac{(\alpha ^2 - \gamma ^2)(\gamma r + 1)}{2\alpha ^2}
- \frac{\gamma ^2}{\alpha ^2} - \frac{3\gamma }{\alpha ^2r} -
\frac{3}{\alpha ^2r^2}} \right)e^{ - \gamma r}} \right], \\
 \end{array}} \right.
\end{equation}

де асимптотика відношення D- та S- хвиль задається співвідношенням

\[
\eta = \mathop {\lim }\limits_{r \to \infty } \left[
{\frac{w(r)}{u(r)}} \right] = \frac{\alpha ^2(\beta ^2 - \alpha
^2)t}{(\gamma ^2 - \alpha ^2)^2}.
\]

Функція $w(r)$ поблизу початку координат пропорційна $r^{2}$:

\[
\mathop {\lim }\limits_{r \to 0} w(r) = \frac{\eta (\gamma ^2 -
\alpha ^2)^2}{8\alpha ^2}r^2.
\]

Параметри аналітичних форм (1)-(3) безпосередньо визначаються при
виборі потенціалу взаємодії.

4) Розроблена Парижською групою \cite{Lacombe1981} аналітична
форма для власного потенціалу залишається на даний час найбільш
вживаною:

\begin{equation}
\label{eq4} \left\{ {\begin{array}{l}
 u\left( r \right) = \sum\limits_{j = 1}^N {C_j \exp \left( { - m_j r}
\right),} \\
 w\left( r \right) = \sum\limits_{j = 1}^N {D_j \exp \left( { - m_j r}
\right)\left[ {1 + \frac{3}{m_j r} + \frac{3}{\left( {m_j r}
\right)^2}}
\right],} \\
 \end{array}} \right.
\end{equation}

де $m_j = \beta + (j - 1)m_0 $, $\beta = \sqrt {ME_d } $,
$m_{0}$=0.9 fm$^{ - 1}$. $M$ - нуклонна маса, $E_{d}$ - енергія
зв'язку дейтрона. Пошук коефіцієнтів аналітичної форми (\ref{eq4})
був здійснений для Парижського \cite{Lacombe1981} і Боннського
\cite{Machleidt2001} потенціалів, причому $N$=13 і 11 відповідно.
Асимптотики при $r \to 0$ вибирались як

\[
u \left( r \right) \to r, \quad w \left( r \right) \to r^3.
\]

5) Аналітична форма Дубовіченко \cite{Dubovichenko2000}

\begin{equation}
\label{eq5} \left\{ {\begin{array}{l}
 u(r) = \sum\limits_{i = 1}^N {A_i \exp ( - a_i r^2),} \\
 w(r) = r^2\sum\limits_{i = 1}^N {B_i \exp ( - b_i r^2),} \\
 \end{array}} \right.
\end{equation}

була застосована для апроксимації ХФД, отриманих для потенціалів
Неймегенської групи. Причому значення $N$=13.

6) Для пояснення D- стану дейтрона і правильної асимптотичної
поведінки у \cite{Berezhnoy2005} запропонована нерелятивістська
ХФД

\begin{equation}
\label{eq6} \left\{ {\begin{array}{l}
 u(r) = \frac{N}{\sqrt {4\pi } }\sum\limits_{k = 1}^{n_u } {C_k e^{ - \alpha
_k r}} , \\
 u(r) = \frac{N}{\sqrt {4\pi } }\rho \sum\limits_{k = 1}^{n_w } {D_k e^{ -
\beta _k r}} \left( {1 + \frac{3}{\beta _k r} + \frac{3}{(\beta _k
r)^2}}
\right), \\
 \end{array}} \right.
\end{equation}

\[
N = \sqrt {\sum\limits_{k,j = 1}^{n_u } {\frac{C_k C_j }{\alpha _k
+ \alpha _j }} + \rho ^2\sum\limits_{k,j = 1}^{n_w } {\frac{D_k
D_j }{\beta _k + \beta _j }} } ,
\]

де $\alpha _{i}$, $\beta _{i}$, $C_{i}$, $D_{i}$, $N$, $\rho $ -
дійсні параметри моделі; $n_u = n_w = 3$. Асимптотики при $r \to
0$ вибирались як $u(r) \to r^2$; $w(r) \to r^3$, а набір
параметрів повинен задовольнити умовам

\[
\sum\limits_k {C_k = 0} ;\sum\limits_k {C_k \alpha _k = 0}
;\sum\limits_k {D_k = 0} ;\sum\limits_k {\frac{D_k }{\beta _k^2 }
= 0} .
\]

7) В роботах \cite{Zhaba1, Zhaba2, Zhaba3} запропоновані нові
аналітичні форми ХФД у вигляді добутку степеневої функції $r^{n}$
на суму експоненціальних членів:

\begin{equation}
\label{eq7} \left\{ {\begin{array}{l}
 u(r) = r^{3 / 2}\sum\limits_{i = 1}^N {A_i \exp ( - a_i r^3),} \\
 w(r) = r\sum\limits_{i = 1}^N {B_i \exp ( - b_i r^3).} \\
 \end{array}} \right.
\end{equation}

При $N$=11 здійснювався пошук показника степеневої функції
$r^{n}$. Оптимальними значеннями виявилися $n$=1.47 і $n$=1.01 для
$u(r)$ і $w(r)$ відповідно. Тобто множники перед сумами в
(\ref{eq7}) можна вибрати як $r^{3/2}$ і $r^{1}$. Точність
параметризації (\ref{eq7}) характеризується величинами
\cite{Machleidt2001}

\begin{equation}
\label{eq8} I_S = \left( {\int\limits_0^\infty {\left[ {u(r) -
u_{table} (r)} \right]^2dr} } \right)^{1 / 2}
\end{equation}

і

\begin{equation}
\label{eq9} I_D = \left( {\int\limits_0^\infty {\left[ {w(r) -
w_{table} (r)} \right]^2dr} } \right)^{1 / 2}.
\end{equation}

Але для оцінки точності параметризації замість (\ref{eq8}) і
(\ref{eq9}) зручніше використовувати величину

\begin{equation}
\label{eq10} \chi ^2 = \frac{\sum\limits_{i = 1}^n {\left( {y_i -
f(x_i ;a_1 ,a_2 ,...,a_p )} \right)^2} }{n - p},
\end{equation}

де $n$ - число точок масиву $y_{i}$ чисельних значень хвильових
функцій дейтрона в координатному представленні; $f$ - апроксимуюча
функція $u$ (або $w)$ згідно формул (\ref{eq7});
$a_{1}$,$a_{2}$,\ldots ,$a_{p}$ - параметри; $p$ - число
параметрів (коефіцієнтів у сумах формул (\ref{eq7})). Отже,
\textit{$\chi $}$^{2}$ визначається не тільки формою апроксимуючої
функції $f$, але і числом вибраних параметрів.

Незважаючи на громіздкі і довготривалі розрахунки і мінімізації
\textit{$\chi $}$^{2}$ (до величин менших за 10$^{-4})$,
доводилося апроксимувати чисельні значення ХФД для потенціалів
Неймегенської групи (NijmI, NijmII, Nijm93, Reid93
\cite{stoks1994}) і потенціалу Argonne v18 \cite{Wiringa1995},
масиви чисел яких становили по 839х4 значень в інтервалі $r$=0-25
fm і 1500х4 значень в інтервалі $r$=0-15 fm відповідно. Значення
коефіцієнтів розкладу $A_{i}$, $a_{i}$, $B_{i}$, $b_{i}$ для
(\ref{eq7}) приведено в \cite{Zhaba2, Zhaba3}. Розраховані
радіальні ХФД $u(r)$ і $w(r)$ по аналітичним формах (\ref{eq7}) в
конфігураційному представленні не містять надлишкових вузлів.

Вибір аналітичних форм залежить від задовільності розрахованих по
цим формам параметрів дейтрона (радіус дейтрона $r_{d}$,
електричний квадрупольний момент $Q_{d}$, магнітний момент, вклад
D- стану, асимптотика D/S- стану та ін.).

\textbf{2. Поляризаційні характеристики дейтрона}

Авторами роботи \cite{levchuk2006} вивчалось незв'язане
фотонародження піона на дейтроні в першій області резонансу.
Неполяризований переріз, асиметрія пучка та векторна і тензорна
асиметрії мішені розраховуються в структурі діаграмного підходу.
Взято до уваги полюсні та однопетлеві діаграми з NN розсіянням в
кінцевому стані. Незважаючи на задовільний опис багатьох доступних
експериментальних даних для неполяризованих повних і
диференціальних перерізів та фотонної (для фотонного пучка)
асиметрії, у деяких випадках наявна суттєва розбіжність між
теорією та експериментом. Для реакцій виду $d$(\textit{$\gamma
$,$\pi $}$^{0})$\textit{np}, $d$(\textit{$\gamma $,$\pi $}$^{ -
})$\textit{pp} була виявлена енергетична залежність
диференціального перерізу ${d\sigma } \mathord{\left/ {\vphantom
{{d\sigma } {d\Omega }}} \right. \kern-\nulldelimiterspace}
{d\Omega }$, фотонної асиметрії $\Sigma $ та тензорної асиметрії
мішені $T_{JM}$. Фотонна асиметрія

\begin{equation}
\label{eq11} \Sigma = \frac {
 \left( {\frac{d\sigma }{d\Omega _\pi
}} \right)^{|} - \left( {\frac{d\sigma }{d\Omega _\pi }} \right)^
\bot } {
 \left( {\frac{d\sigma }{d\Omega _\pi }}
 \right)^{|}
 +
  \left( {\frac{d\sigma }{d\Omega _\pi }} \right)^ \bot
},
\end{equation}

де $\left( {\frac{d\sigma }{d\Omega _\pi }} \right)^{|}$, $\left(
{\frac{d\sigma }{d\Omega _\pi }} \right)^ \bot$ - інклюзивний
переріз для фотонів відносно поляризованої паралелі
(перпендикуляру) до площини.

У кіральній теорії збурень \cite{Epelbaum2002} було застосовано
потенціали NLO і NNLO до систем з двома, трьома і чотирма
нуклонами. Розглянуті дві версії кірального потенціалу NNLO, які
відрізняються доданком для двопіонного обміну, добре описують
параметри двонуклонної системи. Розраховані характеристики
(диференціальний переріз ${d\sigma } \mathord{\left/ {\vphantom
{{d\sigma } {d\Omega }}} \right. \kern-\nulldelimiterspace}
{d\Omega }$, аналізуюча здатність $A_{y}$ і тензорні аналізуючі
здатності $T_{11}$, $T_{20}$, $T_{22})$ для пружного \textit{nd}-
розсіяння свідчать про їх енергетичну залежність при енергіях 3;
10; 65 МеВ.

У праці \cite{Gilman2002} проводиться детальний огляд
експериментальних даних та теоретичних розрахунків для функцій
електромагнітної структури дейтрона $A$, $B$ та тензорної
поляризації $t_{20}$, отриманих з пружного \textit{ed}- розсіяння
при високих енергіях взаємодії, а також для перерізу та набору
поляризаційних асиметрій, що витягуються з високо енергетичної
фотосинтеграції дейтрона в реакції \textit{$\gamma $(d,n)p}. Для
останнього процесу диференціальний переріз, векторна поляризаційна
асиметрія мішені $T$, індукована поляризація $p_{y}$ та асиметрія
$\Sigma$ проявляють енергетичну залежність при 300 та 450 МеВ.
Аналогічна поведінка присутня для аналізуючих здатностей
$iT_{11}$, $T_{20}$, $T_{21}$, $T_{22}$ для пружного \textit{dp}-
розсіяння при енергіях 70; 100; 135; 200; 250 MeВ/N
\cite{sekiguchi2011}.

Вимірювання поляризаційних характеристик реакції фрагментації
дейтрона $A(d,p)X$ при проміжкових та високих енергіях залишається
одним із основних інструментів для дослідження внутрішньої
структури дейтрона. У рамках моделі однонуклонного обміну
імпульсна залежність компонента тензора чутливості до поляризації
дейтронів $T_{20}$ повністю визначається ХФД в імпульсному
представленні \cite{Karmanov1981}:

\begin{equation}
\label{eq12} T_{20} = \frac{1}{\sqrt 2 }\frac{2\sqrt 2 u(p)w(p) -
w(p)^2}{u(p)^2 + w(p)^2}.
\end{equation}

Тут імпульсні компоненти ХФД представлені згідно перетворення
Ханкеля:

\begin{equation}
\label{eq13} u(p) = \int\limits_0^\infty {u(r)j_0 (pr)dr} ; \quad
w(p) = \int\limits_0^\infty {w(r)j_2 (pr)dr} ,
\end{equation}

де $ j_{0}$\textit{(pr)} і $j_{2}$\textit{(pr)} - функції Бесселя
нульового і другого порядку. Експериментально компонента $T_{20}$
визначається через перерізи, що відповідають проекціям спіну (+,
0, -) початкового дейтрона на вісь \cite{Karmanov1981}:

\begin{equation}
\label{eq14} T_{20} = \frac{1}{\sqrt 2 }\frac{\left(
{\frac{d\sigma }{d\Omega }} \right)_ + + \left( {\frac{d\sigma
}{d\Omega }} \right)_ - - 2\left( {\frac{d\sigma }{d\Omega }}
\right)_0 }{\left( {\frac{d\sigma }{d\Omega }} \right)_ + + \left(
{\frac{d\sigma }{d\Omega }} \right)_ - + \left( {\frac{d\sigma
}{d\Omega }} \right)_0 }.
\end{equation}

Крім компоненти $T_{20}$, за ХФД в імпульсному представленні
визначається так звана поляризаційна передача $K_{0}$:

\begin{equation}
\label{eq15} K_0 = \frac{u(p)^2 - w(p)^2 - u(p)w(p) / \sqrt 2
}{u(p)^2 + w(p)^2}.
\end{equation}

Тензорна аналізуюча здатність $A_{yy}$ і тензор-тензорна передача
поляризації $K_{yy}$ в імпульсному наближенні теоретично
розраховуються по наступних формулах \cite{Ladygin2002}:

\begin{equation}
\label{eq16} A_{yy} = \frac{T_{00}^2 - T_{11}^2 + 4P^2T_{10}^2
}{T_{00}^2 + 2T_{11}^2 + 4P^2T_{10}^2 },
\end{equation}

\begin{equation}
\label{eq17} K_{yy} = \frac{5T_{11}^2 + T_{00}^2 - 8P^2T_{10}^2
}{T_{00}^2 + 2T_{11}^2 + 4P^2T_{10}^2 }
\end{equation}

де $P=$0.4$p$ - введений параметр; $T_{ij}(p/2)$ - амплітуди, які
визначаються сферичним і квадрупольним формфакторами дейтрона
\cite{Ladygin2002}. Експериментально величину тензорної
аналізуючої здатності $A_{yy}$ отримують з чисел протонів $n^{ +
}$, $n^{ - }$ і $n^{0}$, зареєстрованих для різних мод поляризації
пучка після поправки на мертвий час установки \cite{Azhgirey2005}:

\[
A_{yy} = 2\frac{p_z^ - (n^ + / n^0 - 1) - p_z^ + (n^ - / n^0 -
1)}{p_z^ - p_{zz}^ + - p_z^ + p_{zz}^ - }.
\]

У \cite{Zhaba4, Zhaba1} по ХФД (\ref{eq5}) і (\ref{eq7})
відповідно для потенціалу Reid93 розрахована величина $T_{20}$
добре корелює з результатами роботи \cite{Ladygin2004}, а отримані
значення для $A_{yy}$ співрозмірні з результатами \cite{Ladygin04}
для Боннського потенціалу. Слід відмітити, що величини
поляризаційних характеристик $T_{20}$ і $A_{yy}$ для потенціалу
Reid93 в \cite{Zhaba4, Zhaba1} майже співпадають зі значеннями цих
характеристик для цього потенціалу при їх визначенні у роботі
\cite{Haysak2014}, де ХФД в координатному та імпульсному
представленнях також не містять надлишкових вузлів. Відхилення
присутні тільки для проміжку великих значень імпульсу, оскільки в
цій області має вплив саме вибір аналітичної форми для
апроксимації ХФД в координатному представленні.

У загальному сукупність спінових спостережуваних для пружного
\textit{dp}- розсіяння назад можна записати у вигляді
\cite{Igo1988, Ladygin1997}

\begin{equation}
\label{eq18} C_{\alpha ,\lambda ,\beta ,\gamma } =
\frac{Tr(F\sigma _\alpha S_\lambda F^ + \sigma _\beta S_\gamma
)}{Tr(FF^ + )},
\end{equation}

де $F$ - матриця розсіяння, $S$ - спіновий оператор; $\sigma $ -
(2$\times $2) матриця Паулі. Експериментальні значення приведені в
\cite{Igo1988}, а теоретичні розрахунки запропоновано і
обґрунтовано в \cite{Ladygin1997} в рамках алгоритму
модельно-незалежного аналізу пружного \textit{dp}- розсіяння в
колінеарній геометрії. Показано, що вимірювання 10 спостережуваних
спінових поляризацій першого та другого порядку реалізують повну
експериментальну базу для однозначного визначення амплітуд
зворотного пружного \textit{dp}- розсіяння.

Для реакцій типу $\vec {1} + A \to \overrightarrow {\frac{1}{2}} +
B$ (наприклад, для $(d,n)$ або $(d,p))$ в \cite{Ohlsen1972}
запропоновано узагальнений запис поляризаційних характеристик,
використовуючи такі співвідношення між ними:

\begin{equation}
\label{eq19} I(\theta ,\varphi ) = I_0 (\theta )\left( {1 +
\frac{3}{2}\sum\limits_j {p_j A_j (\theta )} +
\frac{1}{3}\sum\limits_{j,k} {p_{jk} A_{jk} (\theta )} } \right),
\end{equation}

\begin{equation}
\label{eq20} p_{l'} I(\theta ,\varphi ) = I_0 (\theta )\left(
{P_{l'} (\theta ) + \frac{3}{2}\sum\limits_j {p_j K_j^{l'} (\theta
)} + \frac{1}{3}\sum\limits_{j,k} {p_{jk} K_{jk}^{l'} (\theta )} }
\right),
\end{equation}

де $p_{l'} $ - компоненти вихідної поляризації; $A_j (\theta )$,
$A_{jk} (\theta )$ - аналізуючі здатності; $P_{l'} (\theta )$ -
вихідна поляризація (для неполяризованого падаючого потоку);
$K_j^{l'} (\theta )$, $K_{jk}^{l'} (\theta )$ - коефіцієнти
передачі (переносу) поляризації.

\textbf{3. Асиметрія тензорної і векторної поляризацій}

Для кількісного розуміння структури дейтрона, його $S$- і $D$-
станів та поляризаційних характеристик розглядаються різні моделі
NN потенціалу. Розподіл заряду дейтрона добре не відомий з
експерименту, оскільки він здійснюється тільки через використання
як вимірювань поляризації, так і неполяризованих пружніх розсіяних
диференціальних перерізів. Однак його можна визначити
\cite{Abbott20001}. Диференціальний переріз пружного розсіяння
неполяризованих електронів неполяризованими дейтронами без
вимірювання поляризації відбитих електронів і дейтронів задається
формулою \cite{Gilman2002, Donnelly1986, Haftel1980, Azzam2005,
darwish2009}

\begin{equation}
\label{eq21} \frac{d\sigma _0 }{d\Omega _e } = \left(
{\frac{d\sigma }{d\Omega }} \right)_{Mott} \frac{E^ / }{E}\left[
{A(p^2) + B(p^2)tg^2\left( {\frac{\theta _e }{2}} \right)}
\right],
\end{equation}

де \textit{$\theta $}$_{e}$ - кут розсіяння у лабораторній
системі; $p$ - імпульс дейтрона в одиницях fm$^{ - 1}$; $p^2 =
4EE^ / \sin ^2\left( {\frac{\theta _e }{2}} \right)$ - квадрат 4-
імпульсу передачі; $E$ і $E^{ / }$ - початкова і кінцева енергії;
$A(p)$ і $B(p)$ - функції електричної та магнітної структури

\begin{equation}
\label{eq22} A(p^2) = F_C^2 + \frac{8}{9}\eta ^2F_Q^2 +
\frac{2}{3}\eta F_M^2 ;
\end{equation}

\begin{equation}
\label{eq23} B(p^2) = \frac{4}{3}\eta \left( {1 + \eta }
\right)F_M^2 ,
\end{equation}

де $\eta = \frac{p^2}{4M_D^2 }$; $M_{D}$=1875,63 МеВ - дейтронна
маса. Тут зарядовий $F_{C}(p)$, квадрупольний $F_{Q}(p)$ і
магнітний $F_{M}(p)$ формфактори містять інформацію про
електромагнітні властивості дейтрона:

\begin{equation}
\label{eq24} F_C = \left[ {G_{Ep} + G_{En} }
\right]\int\limits_0^\infty {\left[ {u^2 + w^2} \right]j_0 dr} ;
\end{equation}

\begin{equation}
\label{eq25} F_Q = \frac{2}{\eta }\sqrt {\frac{9}{8}} \left[
{G_{Ep} + G_{En} } \right]\int\limits_0^\infty {\left[ {uw -
\frac{w^2}{\sqrt 8 }} \right]j_2 dr}
\end{equation}

\begin{equation}
\label{eq26}
\begin{array}{l}
 F_M = 2\left[ {G_{Mp} + G_{Mn} } \right]\int\limits_0^\infty {\left[
{\left( {u^2 - \frac{w^2}{2}} \right)j_0 + \left( {\frac{uw}{\sqrt
2 } +
\frac{w^2}{2}} \right)j_2 } \right]dr + } \\
 + \frac{3}{2}\left[ {G_{Ep} + G_{En} } \right]\int\limits_0^\infty
{w^2\left[ {j_0 + j_2 } \right]dr} , \\
 \end{array}
\end{equation}

\noindent де $u$ і $w$ - радіальні ХФД в координатному
представленні; $j_{0}$, $j_{2}$ - сферичні функції Бесселя від
аргументу \textit{pr}/2; $G_{Ep}$ і $G_{En}$ - протонний і
нейтронний електричний формфактори; $G_{Mp}$ і $G_{Mn}$ -
протонний і нейтронний магнітний формфактори. Ці формфактори рівні

\[
G_{Ep} = \left( {1 + \frac{p^2}{18.235}} \right)^{ - 2}; \quad
G_{En} = 0; \quad G_{Mp} = \mu _p G_{Ep} ; \quad G_{Mn} = \mu _n
G_{Ep} ;
\]

де \textit{$\mu $}$_{p}$ і \textit{$\mu $}$_{n}$ - протонний і
нейтронний магнітний моменти в ядерних магнетонах.

Диференціальний переріз для пружного розсіяння поляризованого
електронного пучка від поляризованої дейтронної мішені задається
виразом в лабораторній системі \cite{Donnelly1986}

\begin{equation}
\label{eq27} \frac{d\sigma }{d\Omega _e }\left( {h;p_z ,p_{zz} }
\right) = \Sigma (\theta ,\varphi ) + h\Delta (\theta ,\varphi ),
\end{equation}

де $h$ - спіральність падаючого електронного пучка;$ p_{z}$ і
$p_{zz}$ визначають степінь векторної та тензорної поляризацій
дейтронної мішені. Напрямок поляризації дейтрона визначається
кутами \textit{$\theta $} та \textit{$\phi $} у системі, де вісь
$z$ знаходиться вздовж напрямку віртуального фотона, а вісь $y$
визначається векторами напрямків руху вхідного та вихідного
електронів (Рис. 1).

\pdfximage width 135mm {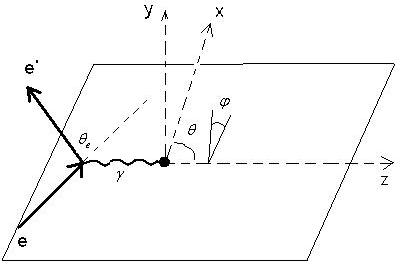}\pdfrefximage\pdflastximage

Fig.~1. Кінематика і системи координат для розсіяння поляризованих
електронів від поляризованої дейтронної мішені

Перший доданок правої частини формули (\ref{eq27}) задає переріз
для неполяризованого електрона

\begin{equation}
\label{eq28} \Sigma (\theta ,\varphi ) = \frac{d\sigma _0
}{d\Omega _e }\left[ {1 + \Gamma (\theta ,\varphi )} \right],
\end{equation}

де $\frac{d\sigma _0 }{d\Omega _e }$ - неполяризований
диференційний переріз. Величина $\Gamma $ описує поляризовану
дейтронну мішень і містить тензорні дейтронні аналізуючі здатності
$T_{2j}$:

\begin{equation}
\label{eq29}
\begin{array}{l}
 \Gamma (\theta ,\varphi ) = p_{zz} \left[ {\frac{1}{\sqrt 2 }} \right.P_2^0
(\cos \theta )T_{20} (p^2,\theta _e ) - \frac{1}{\sqrt 3 }P_2^1
(\cos \theta
)\cos \varphi T_{21} (p^2,\theta _e ) + \\
 + \left. {\frac{1}{2\sqrt 3 }P_2^2 (\cos \theta )\cos (2\varphi )T_{22}
(p^2,\theta _e )} \right]. \\
 \end{array}
\end{equation}

Другий доданок правої частини формули (\ref{eq27}) описує
спірально-залежний диференціальний переріз для поляризованого
пучка електронів та поляризованої дейтронної мішені і містить
векторні дейтронні аналізуючі здатності $T_{10}$ і $T_{11}$:

\begin{equation}
\label{eq30} h\Delta (\theta ,\varphi ) = h\frac{d\sigma _0
}{d\Omega }p_z \left[ {\frac{\sqrt 3 }{2}P_1 (\cos \theta )t_{10}
(p^2,\theta _e ) - \sqrt 3 P_1^1 (\cos \theta )\cos \varphi t_{11}
(p^2,\theta _e )} \right].
\end{equation}

В формулах (\ref{eq29}) і (\ref{eq30}) $P_l (x)$ і $P_l^m (x)$ -
поліноми Лежандра і приєднані поліноми Лежандра відповідно.

В експериментах по неполяризованному пружному розсіянні функції
структури можуть бути отриманими, визначаючи $B(p)$ безпосередньо
із поперечного перерізу розсіяння назад. Рівняння (\ref{eq23})
містить магнітний формфактор $F_{M}(p)$, який з $F_{C}(p)$ і
$F_{Q}(p)$ є складовою частиною і в формулі (\ref{eq22}). Отже,
додатком до неполяризованого розсіяння повинна також розглядатися
і помітна поляризація \cite{Azzam2005}. Оскільки спін дейтрона
рівний $S=$1, то тензорна і векторна поляризації є також
помітними, і можуть бити розрахованими і розглядатися в задачах
розсіяння. У свою чергу величини тензорної $T_{20}$, $T_{21}$,$
T_{22}$ і векторної $T_{10}, T_{11}$ дейтронних аналізуючих
здатностей визначаються через формфактори як \cite{darwish2008,
darwish2009} (у еквівалентних більш широковживаних термінах
\cite{arnold1981, Abbott20001, Gilman2002} - це тензорна $t_{2j}$
і векторна $t_{1i}$ поляризації):

\begin{equation}
\label{eq31} t_{20} (p,\theta _e ) = - \frac{1}{\sqrt 2 S}\left(
{\frac{8}{3}\eta F_C (p)F_Q (p) + \frac{8}{9}\eta ^2F_Q^2 (p) +
\frac{1}{3}\eta \left[ {1 + 2(1 + \eta )tg^2\left( {\frac{\theta
_e }{2}} \right)} \right]F_M^2 (p)} \right),
\end{equation}

\begin{equation}
\label{eq32} t_{21} (p,\theta _e ) = \frac{2}{\sqrt 3 S\cos \left(
{\frac{\theta _e }{2}} \right)}\eta \sqrt {\eta + \eta ^2\sin
^2\left( {\frac{\theta _e }{2}} \right)} F_M (p)F_Q (p),
\end{equation}

\begin{equation}
\label{eq33} t_{22} (p,\theta _e ) = - \frac{1}{2\sqrt 3 S}\eta
F_M^2 (p),
\end{equation}

\begin{equation}
\label{eq34} t_{11} (p,\theta _e ) = \frac{2}{\sqrt 3 S}\sqrt
{\eta (1 + \eta )} F_M (p)\left[ {F_C (p) + \frac{\eta }{3}F_Q
(p)} \right]tg\left( {\frac{\theta _e }{2}} \right),
\end{equation}

\begin{equation}
\label{eq35} t_{10} (p,\theta _e ) = - \sqrt {\frac{2}{3}}
\frac{\eta }{S}\sqrt {(1 + \eta )\left( {1 + \eta \sin ^2\left(
{\frac{\theta _e }{2}} \right)} \right)} F_M^2 (p)tg\left(
{\frac{\theta _e }{2}} \right)\sec \left( {\frac{\theta _e }{2}}
\right),
\end{equation}

де фактор $S(p,\theta _e ) = A(p) + B(p)tg^2\left( {\frac{\theta
_e }{2}} \right)$ визначається функціями структури і кутом
\textit{$\theta $}$_{е}$. Величини тензорної $t_{20}$ і векторної
$t_{11}$ поляризації визначаються формфакторами $F_{C}(p)$,
$F_{Q}(p)$, $F_{M}(p)$ і кутом розсіяння \textit{$\theta $}$_{е}$,
а $t_{21}-F_{Q}(p)$, $F_{M}(p)$ і \textit{$\theta $}. Величини
$t_{22}$ і $t_{10}$ залежить тільки від формфактору $F_{M}(p)$ і
від кута розсіяння.

Поляризація відбитого дейтрона може бути виміряна, якщо детально
аналізувати процес розсіяння. Диференціальний переріз для
подвійного процесу розсіяння \cite{Gilman2002, Arnold1980}

\begin{equation}
\label{eq36}
\begin{array}{l}
 \frac{d\sigma }{d\Omega d\Omega _2 } = \left. {\frac{d\sigma }{d\Omega
d\Omega _2 }} \right|_0 \left[ {1 + \frac{3}{2}ht_{11} A_y \sin
\varphi _2 +
\frac{1}{\sqrt 2 }t_{20} A_{zz} - } \right. \\
 \mbox{ } - \left. {\frac{2}{\sqrt 3 }t_{21} A_{xz} \cos \varphi _2 +
\frac{1}{\sqrt 3 }t_{22} (A_{xx} - A_{yy} )\cos 2\varphi _2 } \right] \\
 \end{array}
\end{equation}

де $h=\pm $1/2 - поляризація падаючого електронного пучка;
\textit{$\phi $}$_{2}$- кут між двома розсіюючими площинами;
$A_{y}$ і $A_{ij}$ - векторна і тензорна аналізуючі здатності
вторинного розсіяння; $t_{20}$, $t_{21}$ і $t_{22}$ визначаються
формулами (\ref{eq31})-(\ref{eq33}).

У роботі \cite{darwish2009} представлені результати для тензорної
поляризації $t_{20}$, яка залежить від імпульсу $р$ та кута
розсіяння електронів \textit{$\theta $}$_{e}$. Проілюстрована
асиметрія для $t_{20}$ у залежності від кута \textit{$\theta
$}$_{e}$. Показано, що асиметрія $t_{20}$ майже незалежна від
вільних нуклонних формфакторів і, зокрема, від маловідомого
нейтронного електричного формфактора. Також знайдено, що величина
$t_{20}$ слабо залежить від кута розсіяння до \textit{$\theta
$}$_{e} \approx $120$^{0}$, оскільки її значення майже ті самі в
цій області. Це чітко випливає з рівняння (\ref{eq31}), тому що
$t_{20}$ прямує до константи $ - 1 \mathord{\left/ {\vphantom {1
{\sqrt 8 }}} \right. \kern-\nulldelimiterspace} {\sqrt 8 }$.

\pdfximage width 135mm {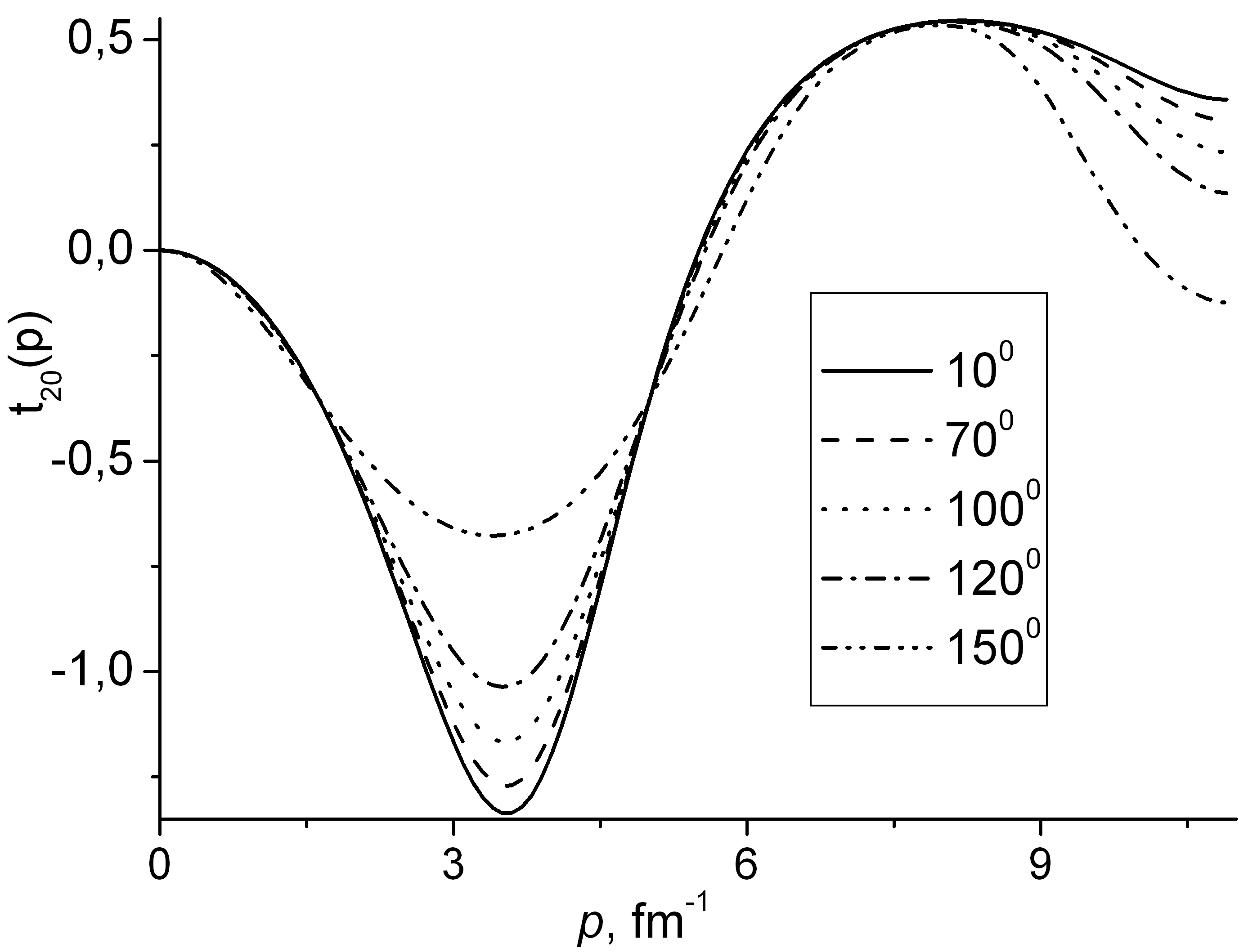}\pdfrefximage\pdflastximage

Fig.~2. Кутова асиметрія тензорної поляризації $t_{20}$

\pdfximage width 135mm {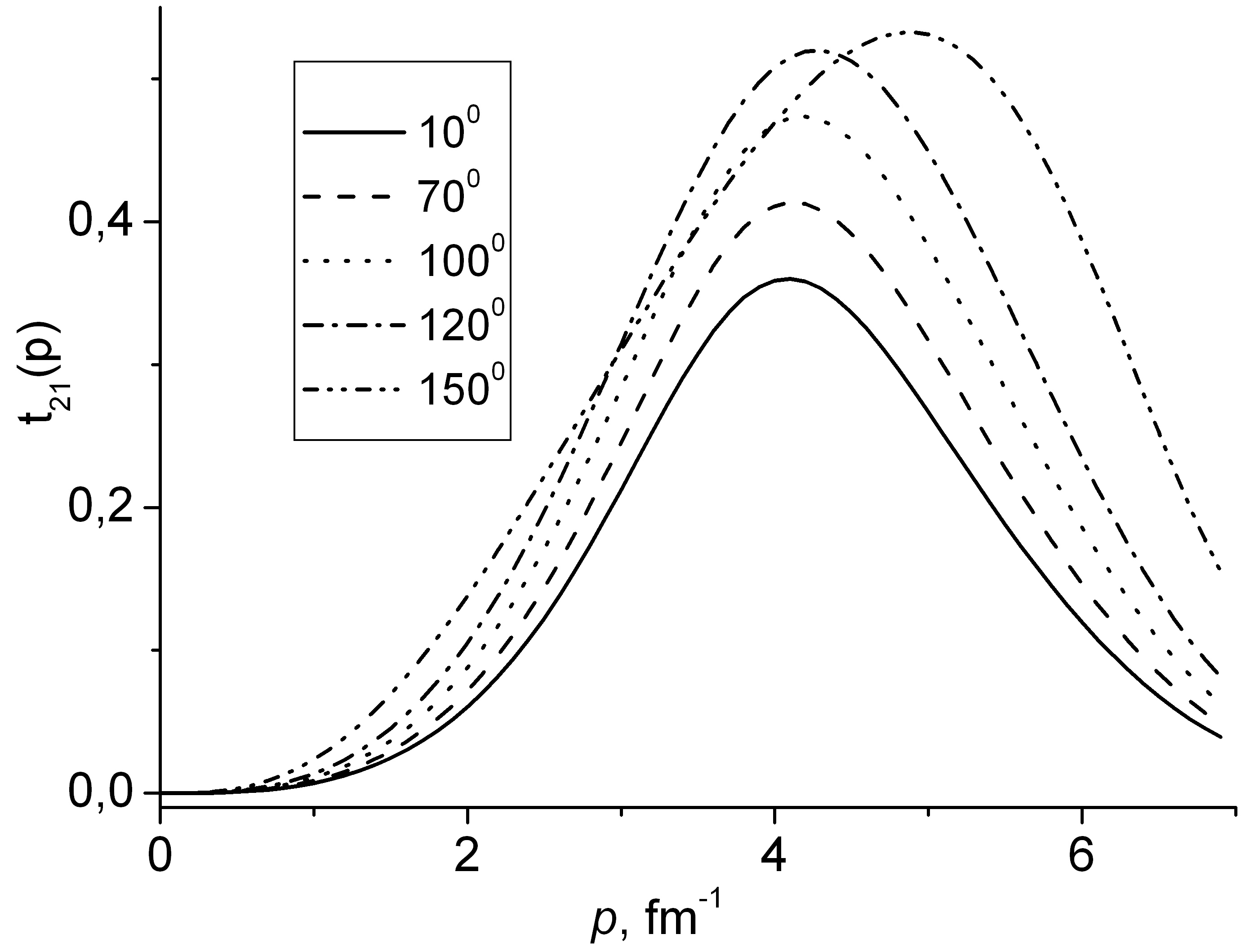}\pdfrefximage\pdflastximage

Fig.~3. Кутова асиметрія тензорної поляризації $t_{21}$

Результати кутової асиметрії для векторних $t_{1i}$ і тензорних
$t_{2j}$ дейтронних поляризацій (\ref{eq31})-(\ref{eq35})
представлені на Рис. 2 - Рис. 6.

\pdfximage width 135mm {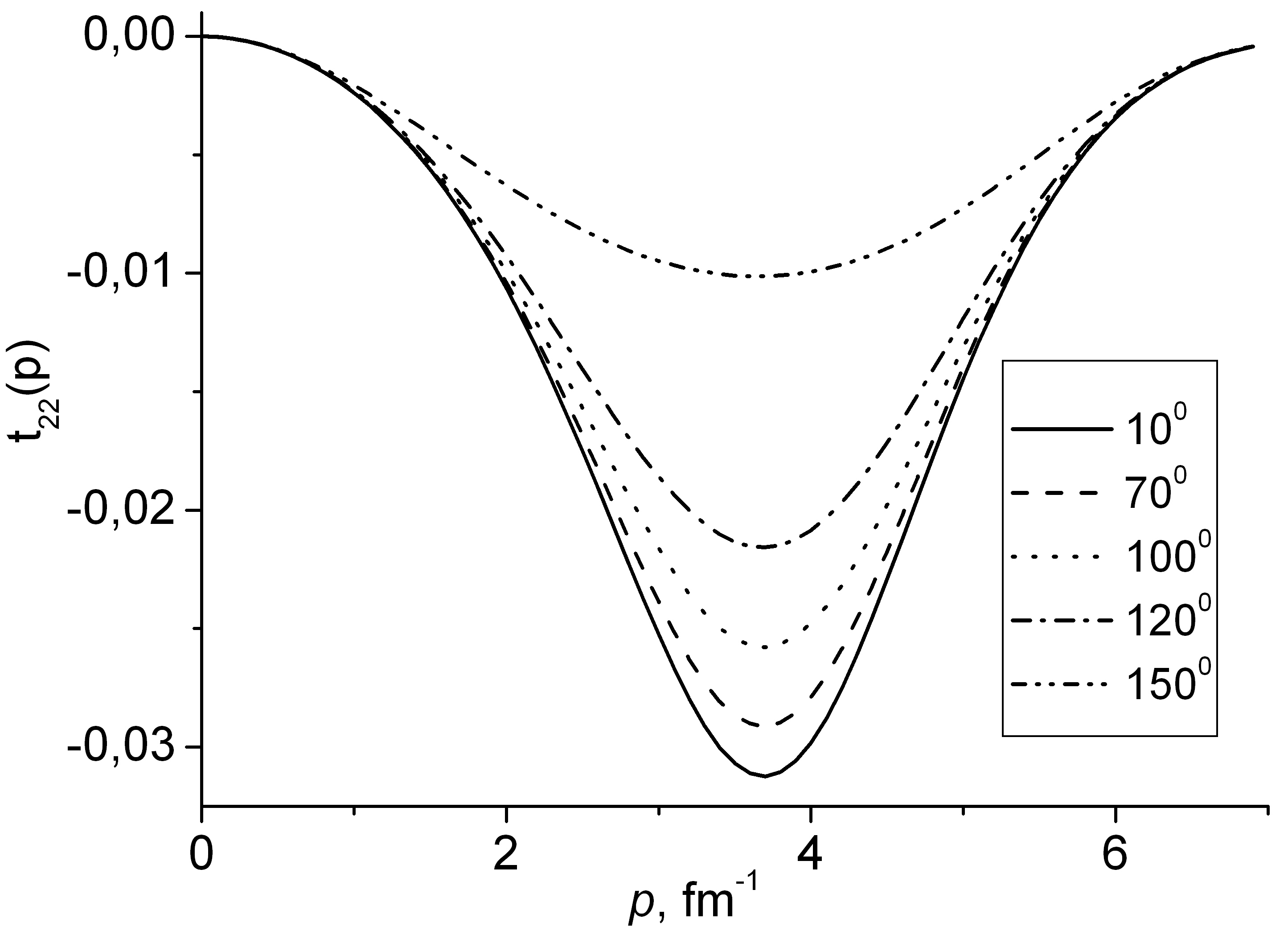}\pdfrefximage\pdflastximage

Fig.~4. Кутова асиметрія тензорної поляризації $t_{22}$

\pdfximage width 135mm {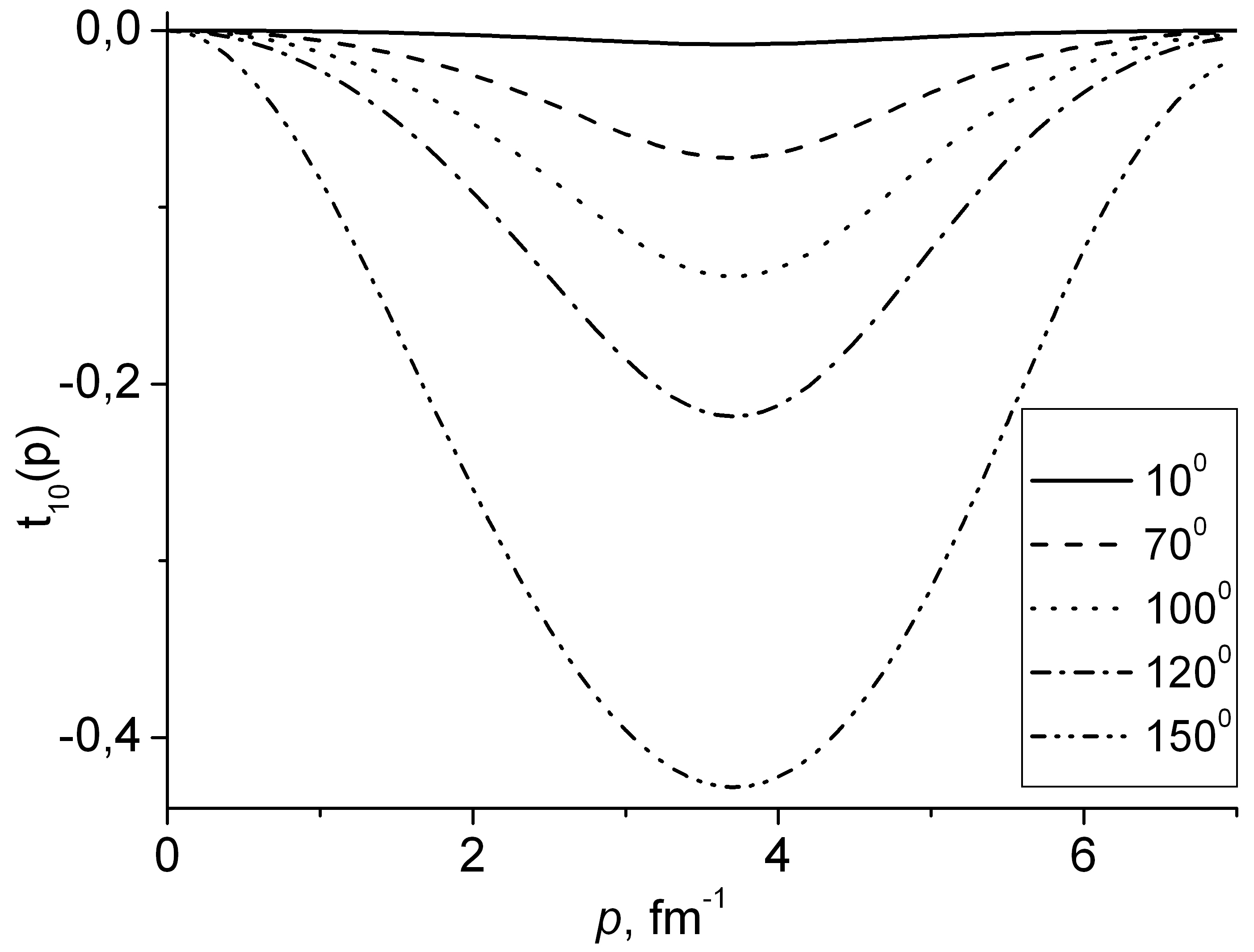}\pdfrefximage\pdflastximage

Fig.~5. Кутова асиметрія векторної поляризації $t_{10}$

Розрахунки проведено по аналітичним формам (\ref{eq7}) з
коефіцієнтами \cite{Zhaba3} для нуклон-нуклонного потенціалу
Reid93.

\pdfximage width 135mm {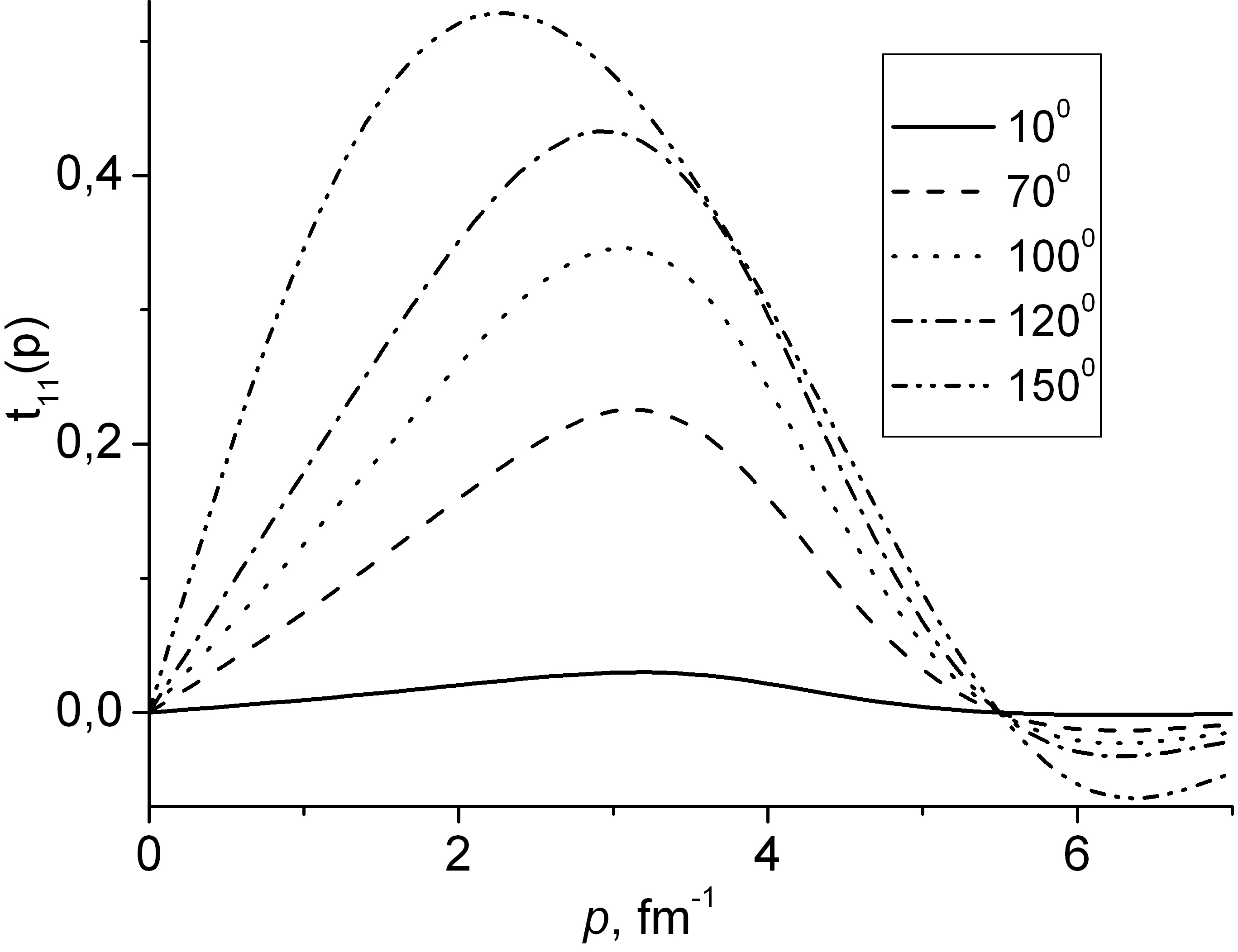}\pdfrefximage\pdflastximage

Fig.~6. Кутова асиметрія векторної поляризації $t_{11}$

Окрім кутової асиметрії, наявна й імпульсна асиметрія для
векторних $t_{1i}$ дейтронних поляризацій (Рис. 7 і 8).

\pdfximage width 135mm {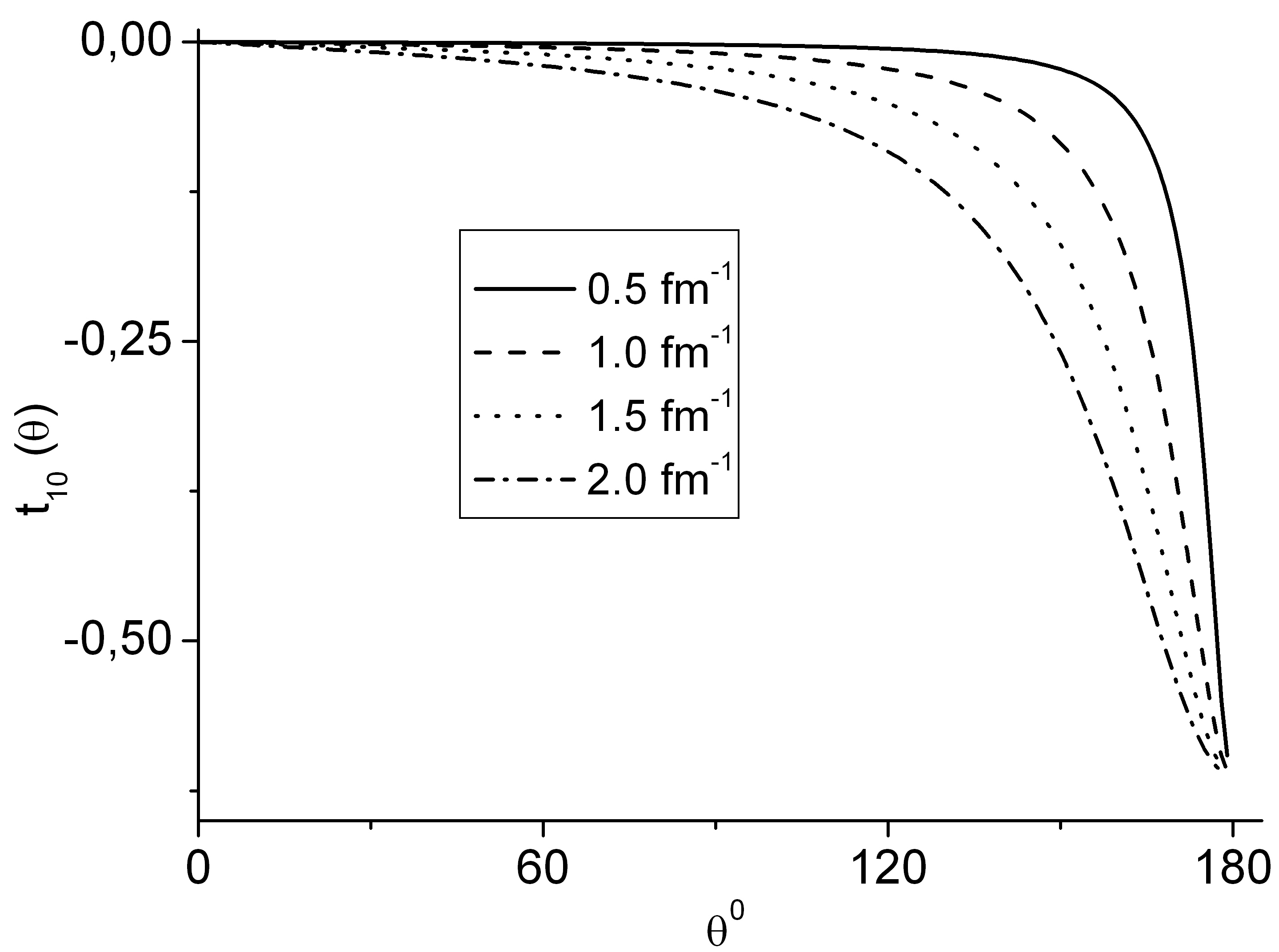}\pdfrefximage\pdflastximage

Fig.~7. Імпульсна асиметрія векторної поляризації $t_{10}$

\pdfximage width 135mm {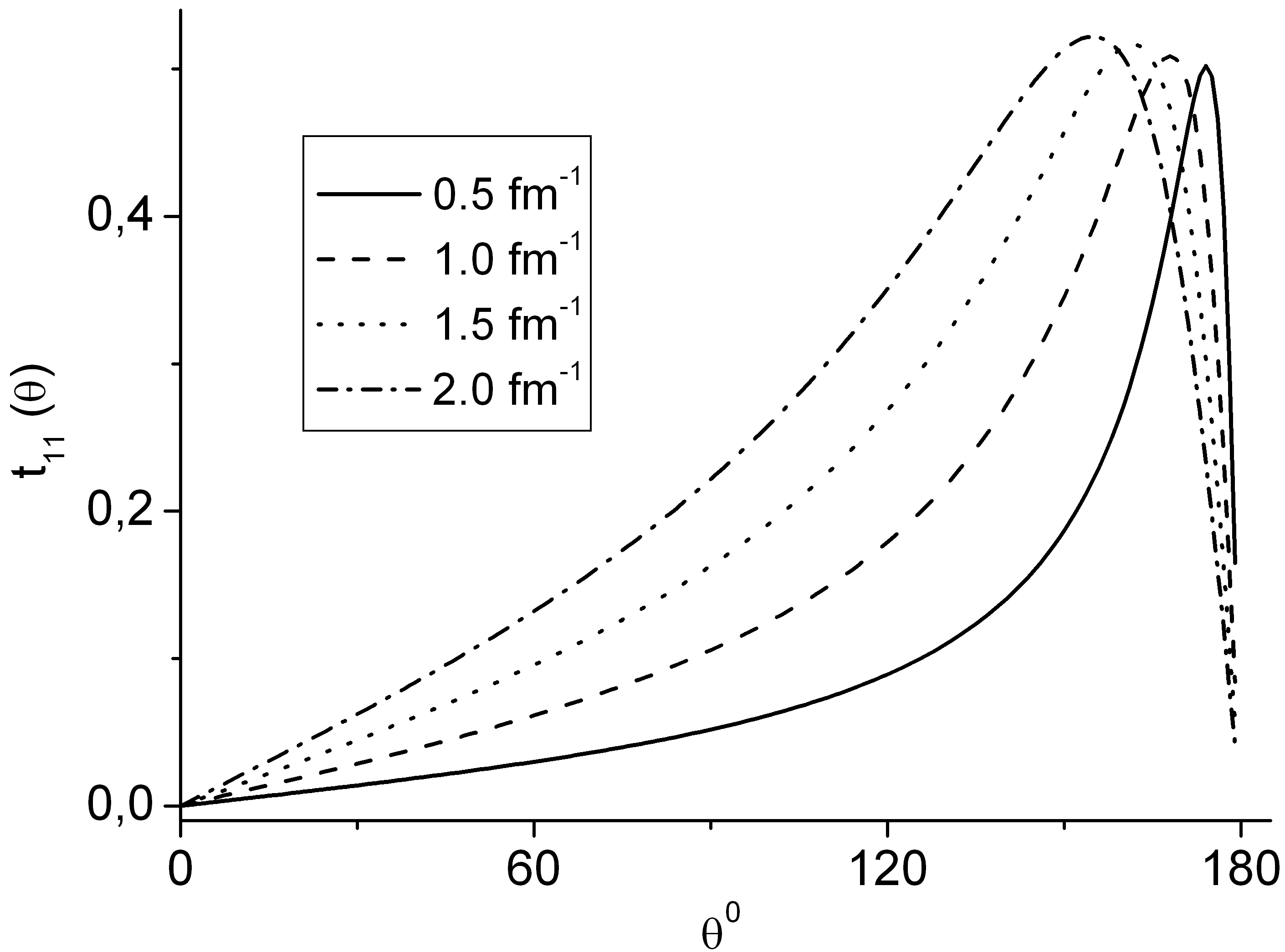}\pdfrefximage\pdflastximage

Fig.~8. Імпульсна асиметрія векторної поляризації $t_{11}$

Слід звернути увагу на те, як саме впливає апроксимація ХФД в
координатному представленні на подальші результати розрахунків
тензорної поляризації $t_{20}$. Так на Рис. 9 приведено порівняння
значень $t_{20}$ при \textit{$\theta $}=70$^{0}$, якщо
застосовувати різні апроксимації ХФД для одного і того ж
нуклон-нуклонного потенціалу Reid93. Використано такі позначення
для ХФД з її коефіцієнтами розкладу: \emph{1} - ХФД з функціями
Лагерра \cite{Zhaba5}; \emph{2} - аналітичні форми Дубовіченко
(\ref{eq5}) \cite{Zhaba4}; \emph{3} - ХФД з \cite{Zhaba6};
\emph{4} - нові аналітичні форми (\ref{eq7}) \cite{Zhaba3}. Крім
цього, порівняно отримані теоретичні значення $t_{20}$ з
експериментальними даними колаборацій Bates \cite{Schulze1984,
The1991, Garcon1994}, BLAST \cite{kohl2008, zhang2011,
hasell2011}, JLab \cite{Abbott20001, kox2001}, NIKHEF
\cite{FerroLuzzi1996, Bouwhuis1999}, Saclay \cite{frois1991},
VEPP-3 \cite{Gilman1990, Nikolenko2001, Nikolenko2003,
nikolenko2010, Zevakov2006} та оглядів Boden \cite{Boden1991},
Garcon \cite{Garcon1994}, Abbott \cite{Abbott2000}. Хороше
узгодження значень наявне для області значень імпульсів
$p$=1-4~fm$^{-1}$. На жаль, при більших імпульсах відсутні
експериментальні дані.

\pdfximage width 135mm {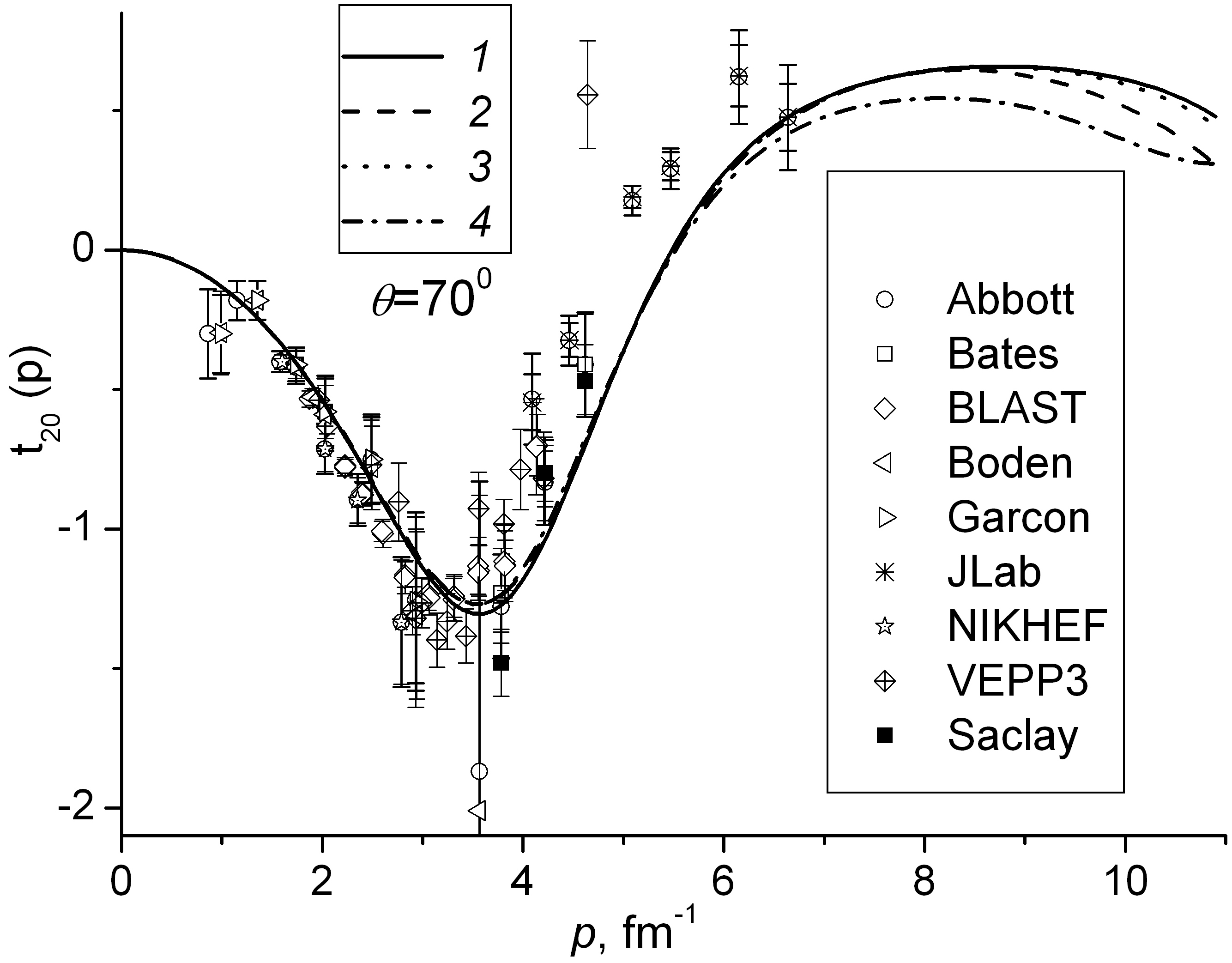}\pdfrefximage\pdflastximage

Fig.~9. Тензорна поляризація $t_{20}$

На Рис. 10 та 11 приведені розраховані величини тензорних
поляризацій$ t_{21}$ і $t_{22}$ (\textit{$\theta $}=70$^{0})$.
Вказано експериментальні дані колаборацій Bates \cite{Garcon1994},
BLAST \cite{zhang2011, hasell2011}, JLab \cite{Abbott20001},
Saclay \cite{hafidi2000}, VEPP-3 \cite{Nikolenko2003,
nikolenko2010, Zevakov2006} для $t_{21}$ і Bates
\cite{Garcon1994}, JLab \cite{Abbott20001}, NIKHEF
\cite{FerroLuzzi1996}, Saclay \cite{hafidi2000} для $t_{22}$. До
речі, наявний значний розкид експериментальних даних. У порівнянні
з $t_{20}$ розраховані величини $t_{21}$ (як і $t_{22})$ при
великих імпульсах співпадають незалежно від вибору ХФД.

\pdfximage width 135mm {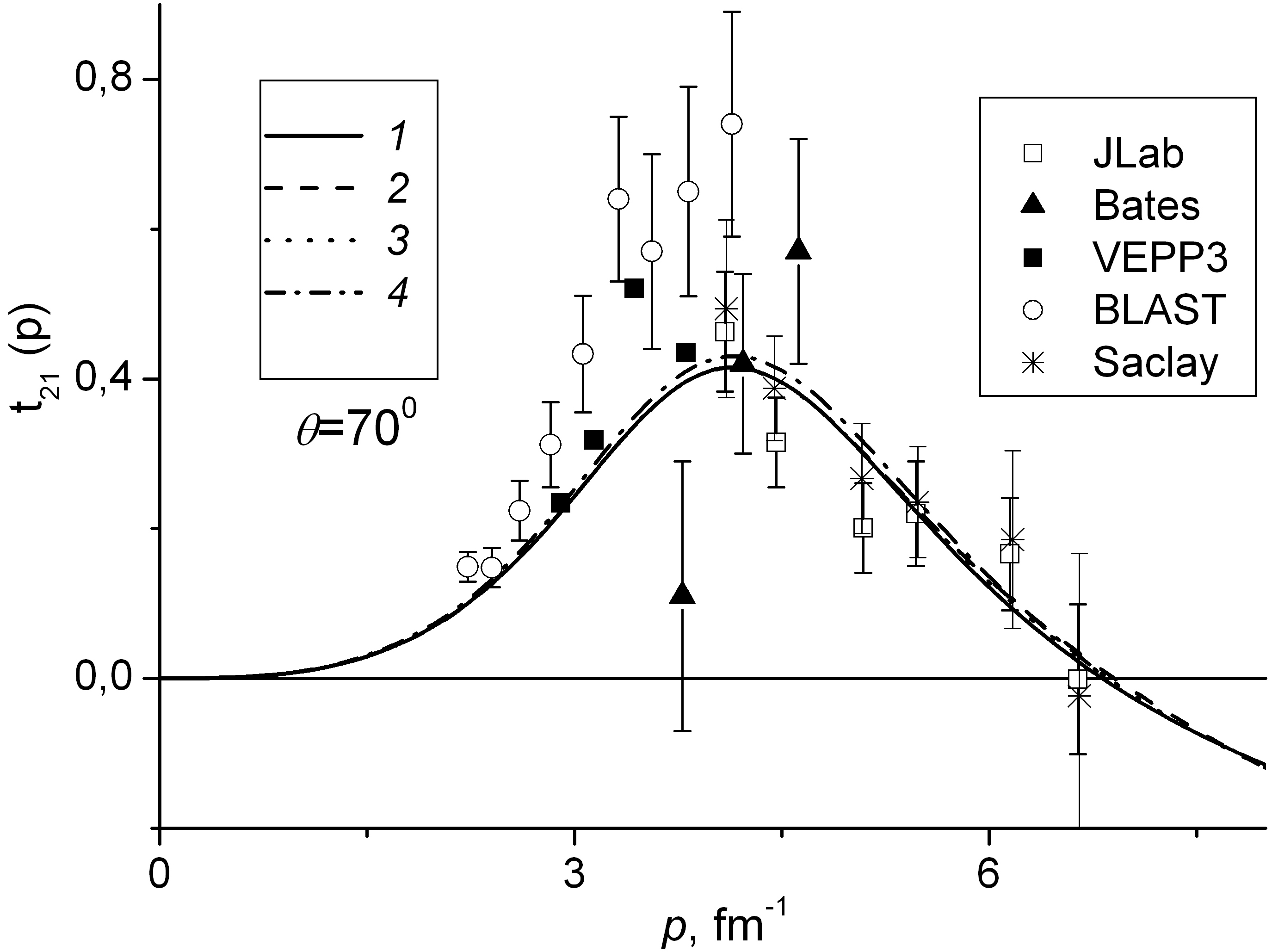}\pdfrefximage\pdflastximage

Fig.~10. Тензорна поляризація $t_{21}$

\pdfximage width 135mm {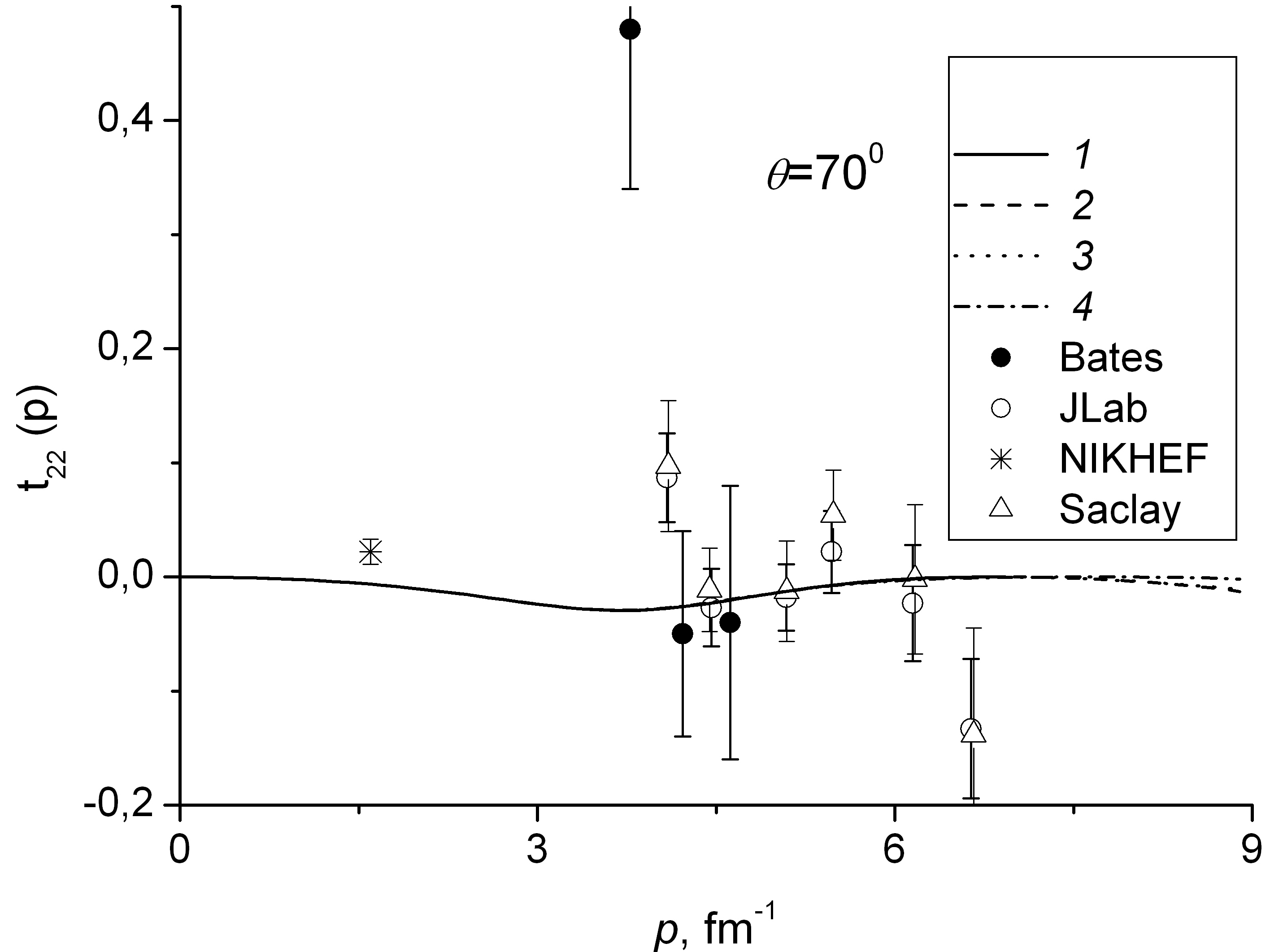}\pdfrefximage\pdflastximage

Fig.~11. Тензорна поляризація $t_{22}$

На Рис. 12 приведені результати розрахунків векторних поляризацій
$t_{1i}$ при \textit{$\theta $}=35$^{0}$. На відміну від тензорної
поляризація її векторна компонента менше досліджена
експериментаторами. Наявні тільки дані для BLAST \cite{Karpius,
hasell2011}. Розраховані значення поляризацій $t_{1i}$ можуть
слугувати певним теоретичним передбаченням для подальших
експериментальних досліджень.

У цілому в науковій літературі відсутні експериментальні дані для
$t_{21}$, $t_{22}$, $t_{10}$ і $t_{11}$ в широкому інтервалі
імпульсів. Тому є актуальним одержання даних величин як
теоретично, так і експериментально.

\pdfximage width 135mm {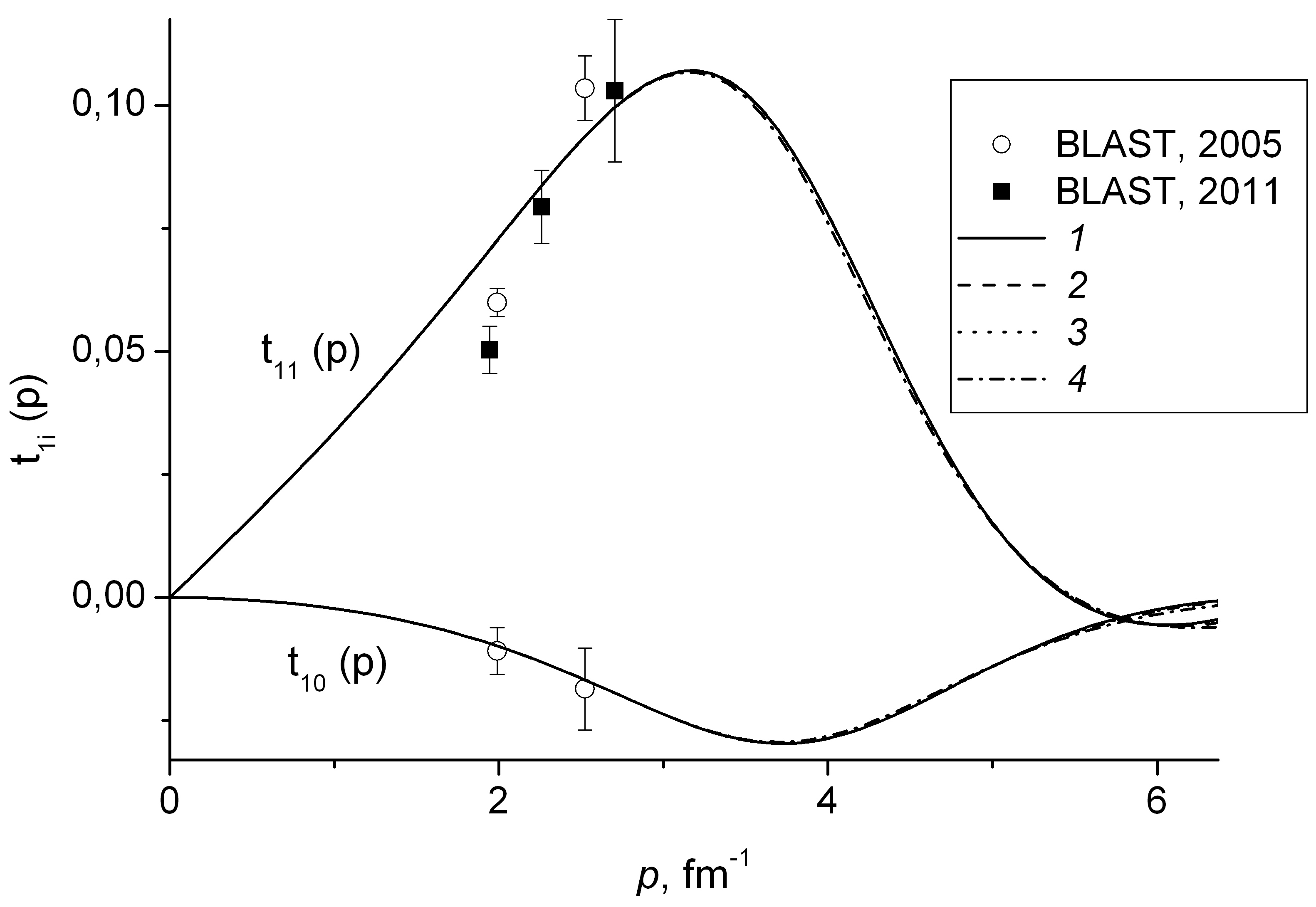}\pdfrefximage\pdflastximage

Fig.~12. Векторні поляризації $t_{10}$ і $t_{11}$

\textbf{4. Асиметрія поляризаційних характеристик
}\textbf{\textit{$\kappa $}}$_{0}$\textbf{,
}\textbf{\textit{T}}$_{2M}$\textbf{ і }\textbf{\textit{R}}

Спінові спостережувані $T_{20},$ \textit{$\kappa $}$_{0}, \quad
K_{xzy}$, $K_{yy}$, $C_{yy}$ в пружному \textit{dp}- розсіянні
назад при проміжкових та високих енергіях були досліджені в
\cite{tanifuji1998} за допомогою методу інваріантної амплітуди з
припущенням однонуклонного обміну. Неузгодженості між теоретичними
розрахунками і експериментальними даними для кореляції
\textit{$\kappa $}$_{0}-T_{20}$ здебільшого знівельовані
включенням ефектів для уявних частин абсорбції в інваріантних
амплітудах. Імпульсна залежність експериментально визначених
$T_{20}$ і \textit{$\kappa $}$_{0}$ для системи протон-нейтрон
пояснена обчисленням із специфікацією ядерних потенціальних
моделей.

Тензорна аналізуюча здатність і поляризаційна передача отримані в
рамках методу інваріантної амплітуди записуються як
\cite{tanifuji1998}

\begin{equation}
\label{eq37} T_{20} = \left\{ {2\sqrt 2 R\cos \Theta - R^2 -
32R'^2 + 12RR'\cos (\Theta ' - \Theta )} \right\} / N_R ,
\end{equation}

\begin{equation}
\label{eq38} \kappa _0 = \left\{ {\sqrt 2 - R\cos \Theta - 4R'\cos
\Theta ' - 3\sqrt 2 RR'\cos (\Theta ' - \Theta ) - 30\sqrt 2 R'^2}
\right\} / N_R ,
\end{equation}

де

\[
N_R = \sqrt 2 + 2\sqrt 2 R^2 + 34\sqrt 2 R'^2 - 4R'\cos \Theta ';
\]

\[
R = \frac{4\left| \rho \right|}{4 + \rho ^2}; \quad R' =
\frac{\rho ^2}{\sqrt 2 (4 + \rho ^2)}; \quad \rho =
\frac{w(p)}{u(p)}.
\]

На Рис. 13 приведено порівняння з експериментальними даними
\cite{punjabi1995} розрахованої величини \textit{$\kappa $}$_{0}$
в залежності від кутів $(\Theta ,\Theta ')$. Наявне добре
співпадання між ними. Кореляція \textit{$\kappa $}$_{0}-T_{20}$
виражається співвідношенням $T_{20} + \sqrt 8 \kappa _0 $ і
вказана на Рис. 14. Аналогічно до попередніх розрахунків
використано ХФД (\ref{eq7}) для потенціалу Reid93.

\pdfximage width 135mm {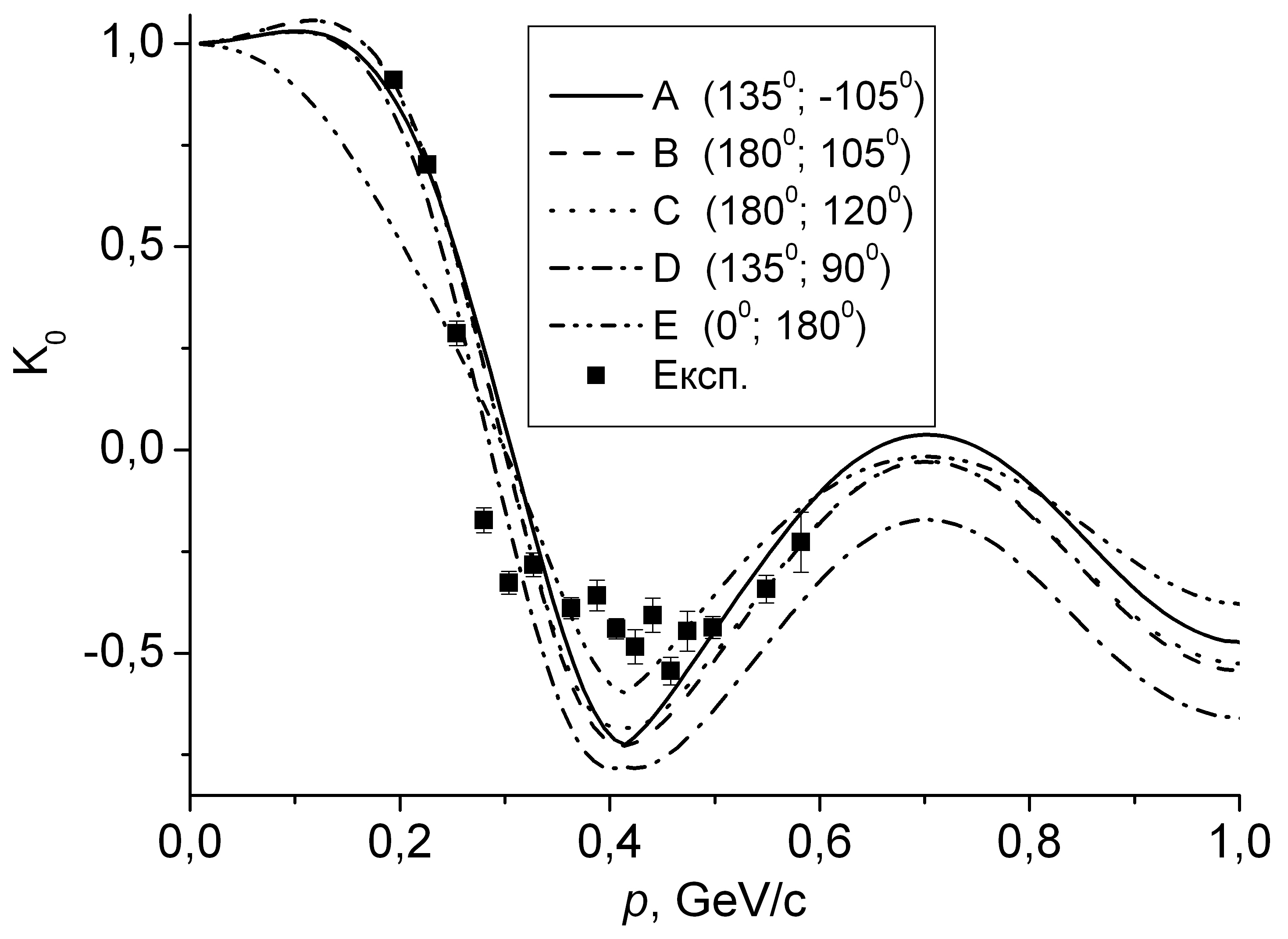}\pdfrefximage\pdflastximage

Fig.~13. Поляризаційна передача \textit{$\kappa $}$_{0}$

\pdfximage width 135mm {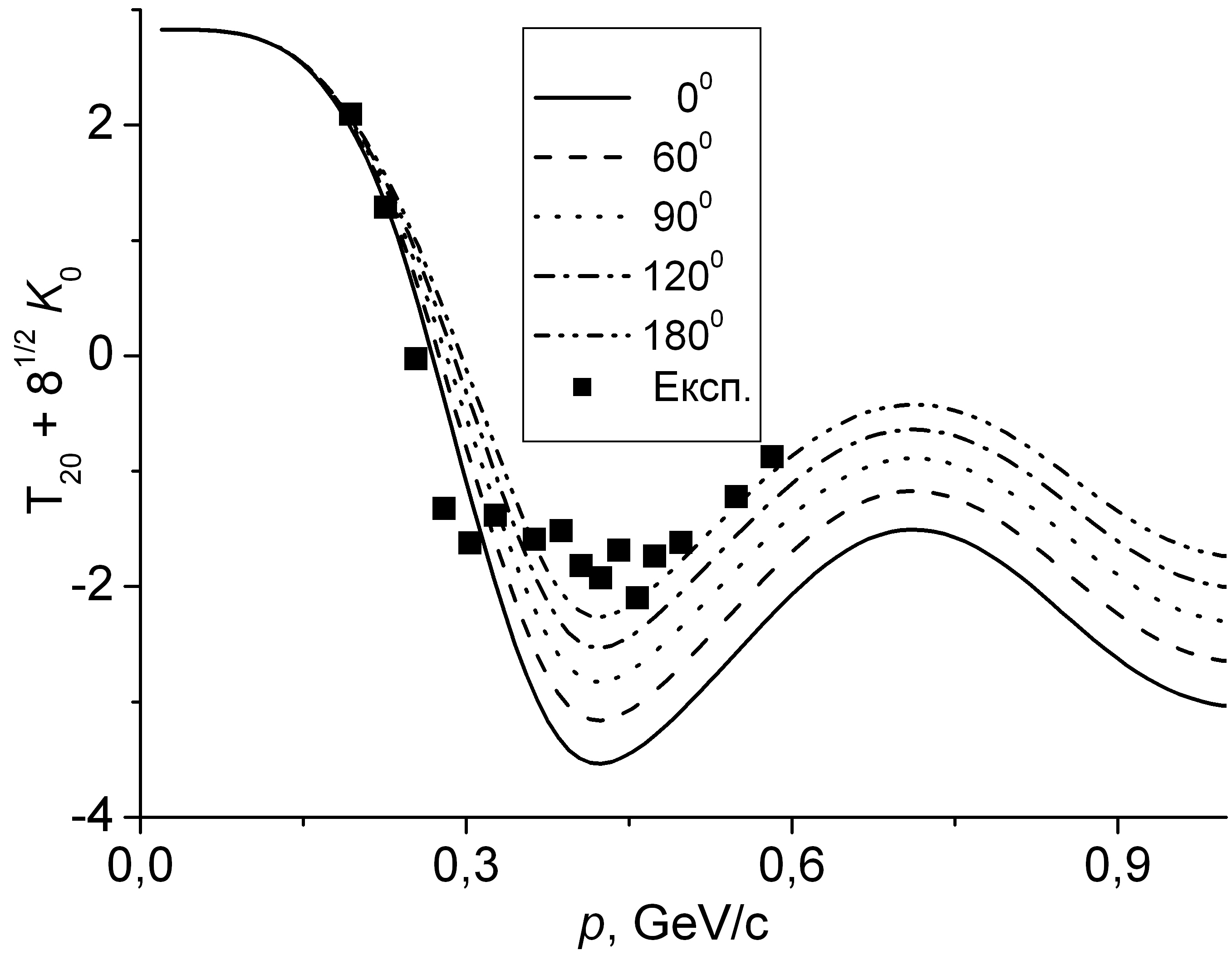}\pdfrefximage\pdflastximage

Fig.~14. Кореляція $T_{20} + \sqrt 8 \kappa _0 $

У роботі \cite{Loginov2004} досліджено компоненти тензорної
аналізуючої здатності в фотонародженні негативного $\pi $- мезона
в реакції за участю дейтрона (процес \textit{$\gamma $d}$ \to \pi
^{ - }$\textit{pp}):

\[
T_{2M} = \frac{SpM\tau _{2M} M^ + }{SpMM^ + },
\]

де \textit{$\tau $}$_{2M}$ - сферичний спін-тензор в дейтронному
матричному розкладі густини. Величина тензорної аналізуючої
здатності $T_{2M}$ визначається за ХФД в імпульсному представленні
і ``протонним'' кутом \textit{$\theta $} \cite{Loginov2004}

\begin{equation}
\label{eq39} T_{20} = \frac{32\sqrt 2 u^2(p) - 16\left( {3\cos
(2\theta ) + 1} \right)u(p)w(p) + \sqrt 2 \left( { - 12\cos
(2\theta ) + 9\cos (4\theta ) + 19} \right)w^2(p)}{4\left[
{16u^2(p) - 4\sqrt 2 \left( {3\cos (2\theta ) + 1} \right)u(p)w(p)
- \left( {6\cos (2\theta ) + 9\cos (4\theta ) - 23} \right)w^2(p)}
\right]};
\end{equation}

\begin{equation}
\label{eq40} T_{21} = 0;
\end{equation}

\begin{equation}
\label{eq41} T_{22} = - \frac{3\sqrt 3 \sin ^2(\theta )w(p)\left(
{4\sqrt 2 u(p) + 3\cos (2\theta ) + 5} \right)w(p)}{16u^2(p) -
4\sqrt 2 \left( {3\cos (2\theta ) + 1} \right)u(p)w(p) - \left(
{6\cos (2\theta ) + 9\cos (4\theta ) - 23} \right)w^2(p)}.
\end{equation}

Якщо кут вилітаючого протона \textit{$\theta $} задовольняє умову
$1 + 3\cos (2\theta ) = 0$ або $3\cos ^2\theta - 1 = 0$, то вирази
для $T_{20}$ і $T_{22}$ спрощуються і набувають наступної форми:

\[
T_{20} = \frac{2u^2(p) + w^2(p)}{2\sqrt 2 \left[ {u^2(p) +
2w^2(p)} \right]};
\]

\[
T_{22} = - \frac{\sqrt 3 w(p)\left[ {\sqrt 2 u(p) + w(p)}
\right]}{2\left[ {u^2(p) + 2w^2(p)} \right]}.
\]

На Рис. 15 та 16 представлена імпульсна асиметрія тензорних
аналізуючих здатностей $T_{20}$ і $T_{22}$. Слід зауважити, що
спостерігається симетрія величин $T_{20}$ і $T_{22}$ відносно кута
90$^{0}$. На відміну від розрахунків в \cite{Loginov2004}
(розрахунки $T_{2і}$ для таких потенціалів як Bonn і Paris та для
параметризації DWF-Certov), то у даній роботі проведено обчислення
для нуклон-нуклонного потенціалу Reid93 в більш широкому діапазоні
імпульсів та кутів \textit{$\theta $}.

\pdfximage width 135mm {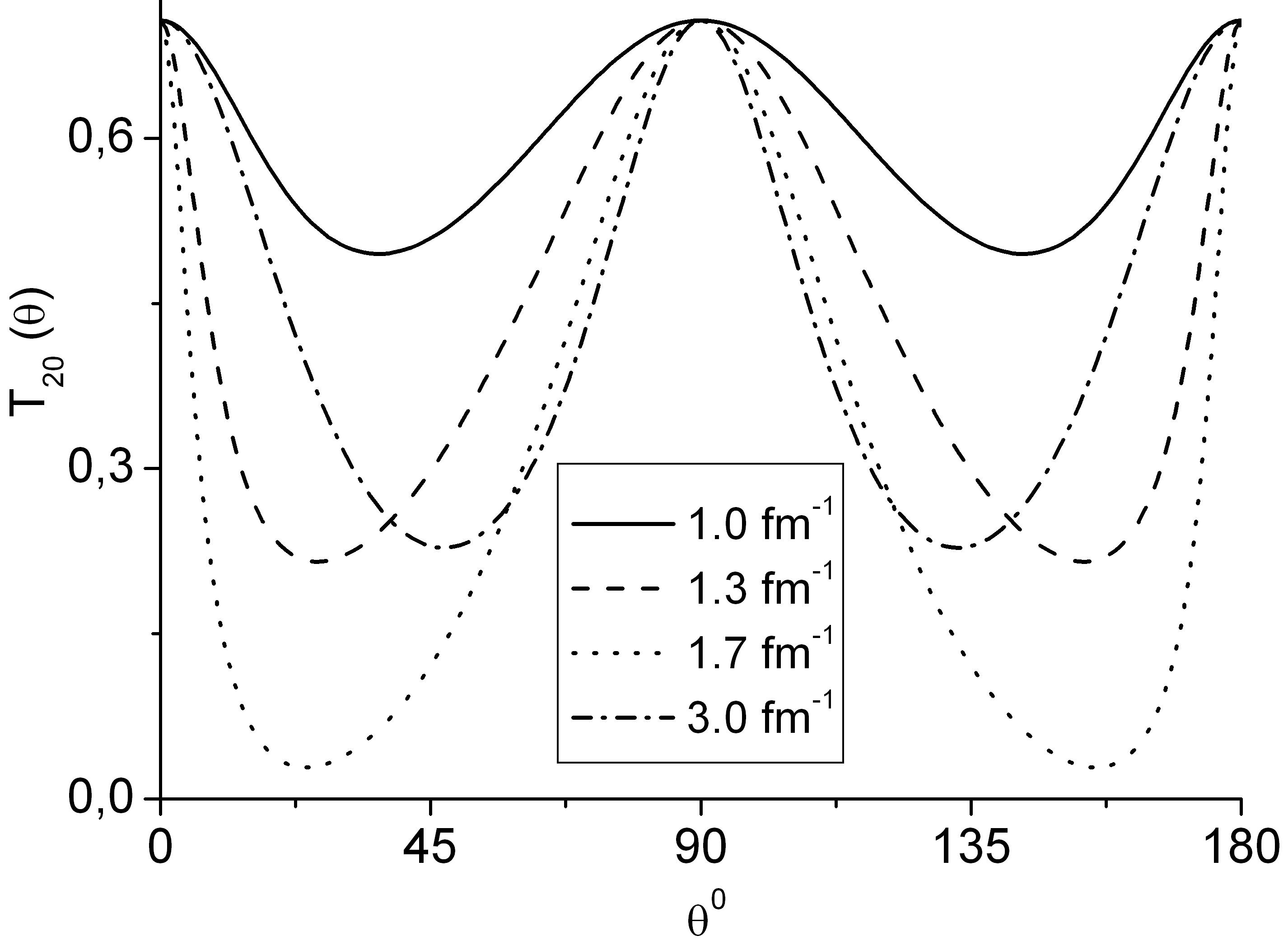}\pdfrefximage\pdflastximage

Fig.~15. Імпульсна асиметрія тензорної аналізуючої здатності
$T_{20}$

\pdfximage width 135mm {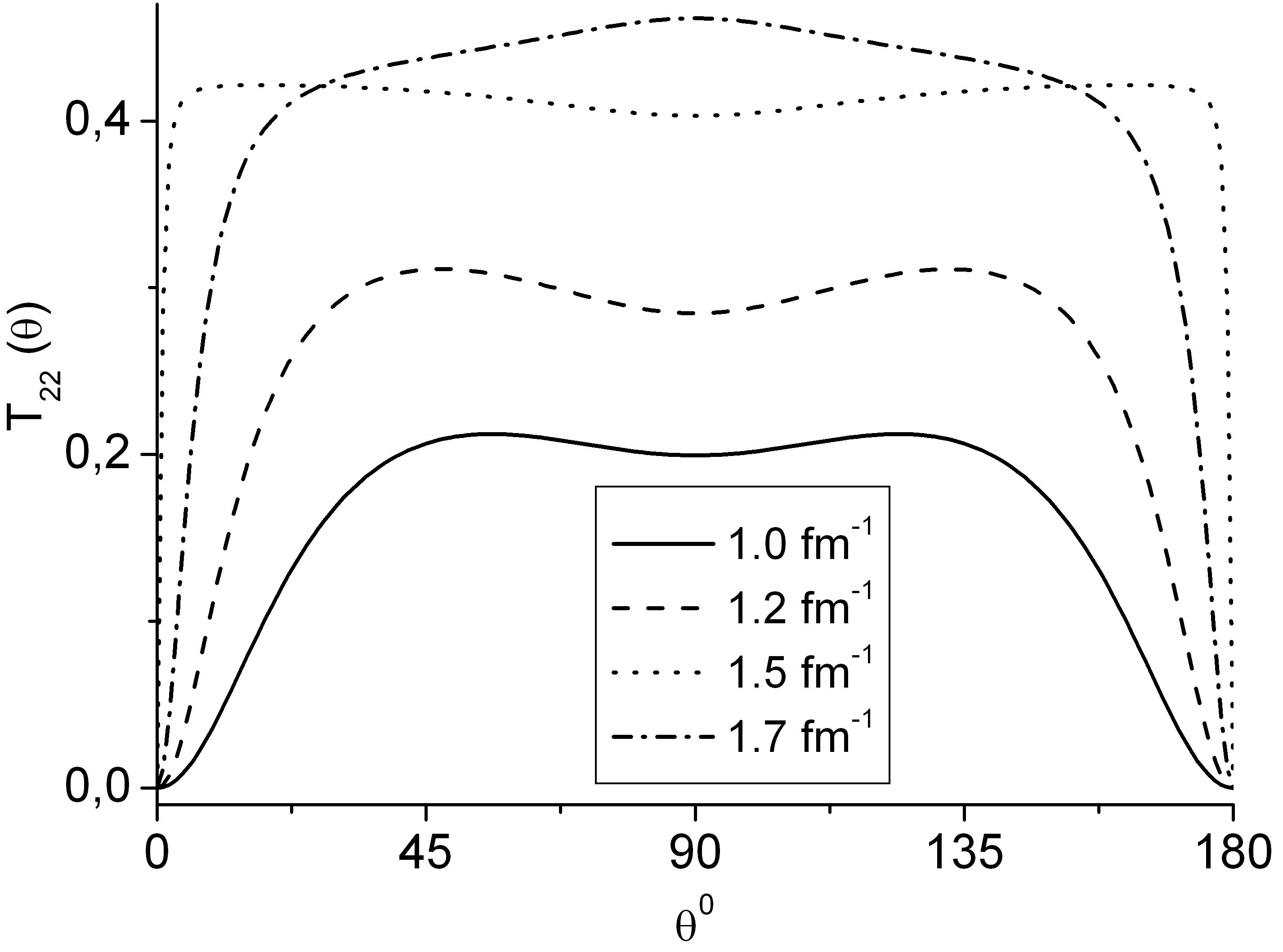}\pdfrefximage\pdflastximage

Fig.~16. Імпульсна асиметрія тензорної аналізуючої здатності
$T_{22}$

Експерименти по дослідженню поляризації дейтронів віддачі націлені
на вимірювання компоненти тензорної поляризації $t_{20}$ і
відношення $R$ \cite{burov1984}. Специфічною особливістю цих
характеристик є те, що вони не залежать від нуклонних
формфакторів, і в межах певних понять і концепцій їх поведінку
визначає функція відносного руху нуклонів. У роботі
\cite{burov1984} показано, що для пружного \textit{ed}- розсіяння
мала 5-8{\%} домішка шести-кваркової компоненти в ХФД зрештою
вносить значну зміну в поведінку функції електричної структури
$A(p^{2})$, тензорної поляризація $T_{20}$ і відношення $R$.
Спостерігається чітка якісна зміна цих поляризаційних
характеристик в області великих імпульсів
25fm$^{-2}$<$p^{2}$<120fm$^{-2}$.

Експериментальне значення тензора поляризованої твердої ND$_{3}$
мішені визначається як \cite{Althoff1990, Boden1991}

\[
R_{EXP} = 1 + P_{zz} (R_T - 1),
\]

де $P_{zz}$ - тензорна поляризація дейтронів мішені. Теоретичні
значення тензора поляризованої мішені і тензорної поляризації
відповідно \cite{Boden1991}

\[
R_T - 1 = \frac{y + 0.5y^2}{1 + 2y^2},
\]

\[
t_{20} = - \sqrt 2 \frac{2y + y^2}{1 + 2y^2},
\]

де $y = \frac{2}{3}\eta \frac{F_Q }{F_C }$. Крім цього, тензорна
поляризація теоретично може бути знайдена як \cite{burov1984}

\[
t_{20} = \frac{1 + x}{\sqrt 2 \left( {1 + x^2 / 8} \right)},
\]

де величина $x$ визначається співвідношенням формфакторів $x =
\sqrt 8 \frac{F_C }{F_Q }$.

Аналогічно до $t_{20}$ відношення $R$ для векторної $P_{x}$ і
тензорної $P_{xz}$ поляризацій задається зручним виразом

\begin{equation}
\label{eq42} R = \frac{P_x }{P_{xz} } = \left[ {2\sin \left(
{\frac{\theta }{2}} \right)\frac{\sqrt {1 + \eta } }{9}\sqrt {1 +
\eta \sin ^2\left( {\frac{\theta }{2}} \right)} } \right]\left[ {1
+ x} \right].
\end{equation}

Кутова асиметрія $R$, розрахована по ХФД (\ref{eq7}) для
потенціалу Reid93, проілюстрована на Рис. 17. Розрахунки проведено
в інтервалі імпульсів до $p$=14 fm$^{-1}$.

\pdfximage width 135mm {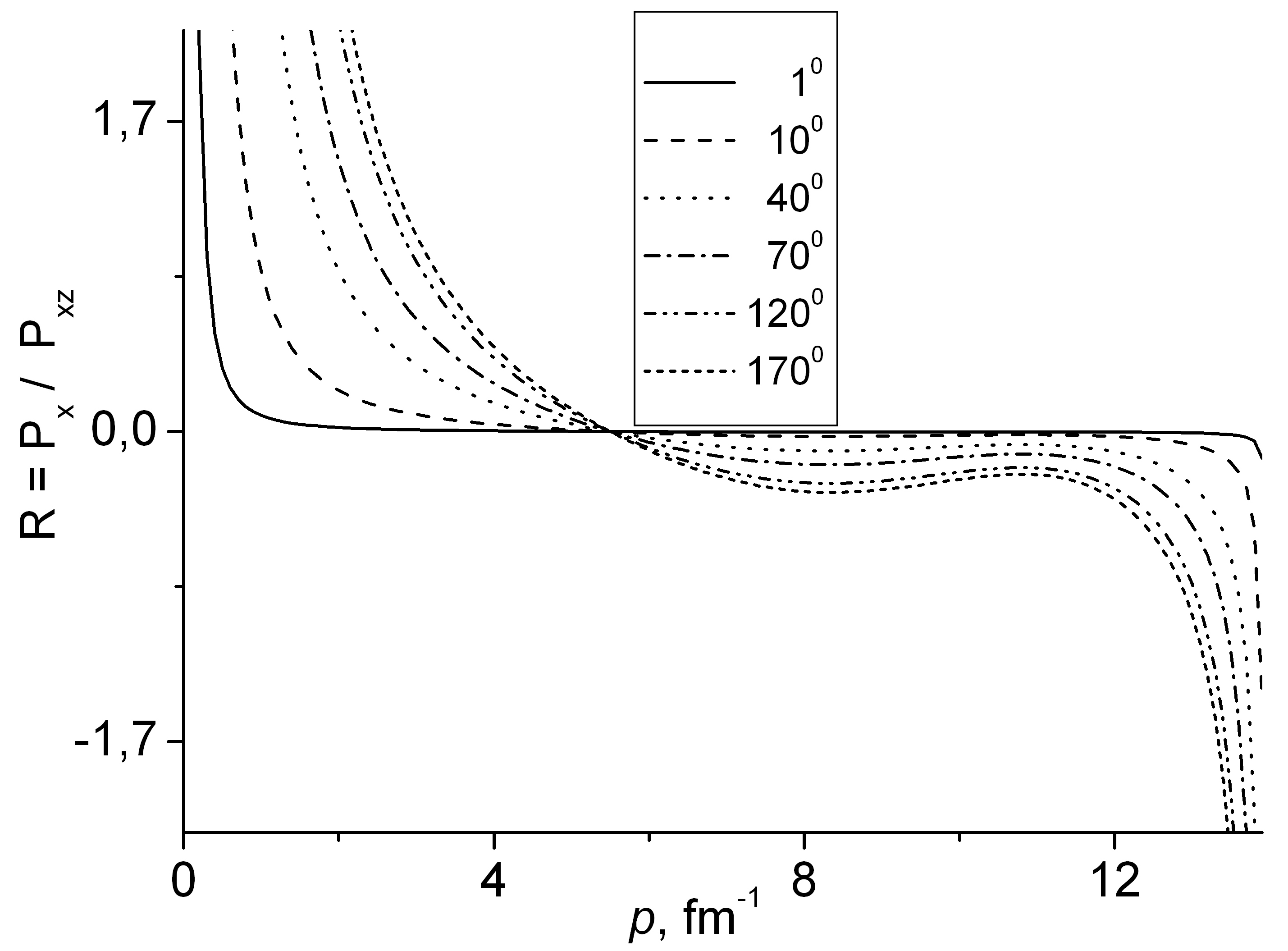}\pdfrefximage\pdflastximage

Fig.~17. Асиметрія $R$

\textbf{Висновки}

Проведено огляд основних аналітичних форм хвильової функції
дейтрона в координатному представленні. Аналізуються асимптотичні
поведінки ХФД поблизу початку координат.

Для розрахунку поляризаційних характеристик дейтрона застосовано
нові аналітичні форми ХФД \cite{Zhaba2, Zhaba3} у вигляді добутку
степеневої функції $r^{n}$ на суму експоненціальних членів
$A_{i}$\textit{$\cdot $exp(-a}$_{i}$\textit{$\cdot $r}$^{3})$. Для
чисельних обчислень використано реалістичний
феноменологічний потенціал Неймегенської групи - Reid93.

У роботі представлені результати кутової асиметрії для векторних
$t_{10}$, $t_{11}$ і тензорних $t_{20}$, $t_{21}$, $t_{22}$
дейтронних поляризацій. Окрім кутової асиметрії, наявна й
імпульсна асиметрія для векторних $t_{1i}$ дейтронних поляризацій.

Досліджено вплив апроксимація ХФД в координатному представленні на
подальші результати розрахунків тензорної поляризації $t_{20}$.
Приведено порівняння значень $t_{20}$ при \textit{$\theta
$}=70$^{0}$, якщо застосовувати чотири різні апроксимації ХФД для
одного і того ж NN потенціалу Reid93. При порівнянні отриманих
теоретичних значень $t_{20}$ з експериментальними даними провідних
колаборацій та оглядів спостерігається хороше узгодження для
області значень імпульсів $p$=1-4 fm$^{-1}$. У зв'язку з
відсутністю експериментальних даних для поляризацій $t_{21}$,
$t_{22}$, $t_{10}$ і $t_{11}$ в широкому інтервалі імпульсів є
актуальним як теоретичне, так і експериментальне одержання даних
величин.

У рамках методу інваріантної амплітуди проведено розрахунок
спінових спостережуваних в пружному \textit{dp}- розсіянні назад -
тензорної аналізуючої здатності $T_{20}$ і поляризаційної передачі
\textit{$\kappa $}$_{0}$. Порівняно значення \textit{$\kappa
$}$_{0}$ та кореляції \textit{$\kappa $}$_{0}-T_{20}$ з
експериментальними даними.

В широкому діапазоні імпульсів та кутів розсіяння \textit{$\theta
$} представлена імпульсна асиметрія тензорних аналізуючих
здатностей $T_{20}$ і $T_{22}$, які характеризують фотонародження
негативного $\pi $- мезона в реакції \textit{$\gamma $(d},$\pi ^{
- })$\textit{pp}. Спостерігається симетрія величин $T_{20}$ і
$T_{22}$ відносно кута 90$^{0}$. Відношення $R$ для векторної
$P_{x}$ і тензорної $P_{xz}$ поляризацій характеризується кутовою
асиметрією.

Отже, поряд з ``$A_{y}$ загадкою'' залишаються актуальними
теоретичні та експериментальні дослідження інших поляризаційних
характеристик для процесів за участю дейтрона, для яких наявна як
кутова, так й імпульсна асиметрія. Отримані результати для
векторної і тензорної дейтронних поляризацій $t_{ij}(p)$ дають
певну інформацію про електромагнітну структуру дейтрона і
диференціальний переріз подвійного розсіяння.

\textbf{Подяки}

Автор висловлює вдячність доц., к.ф.-м.н. Гайсаку І.І.; доц.,
к.х.н. Васильєвій Г.В.; доц., к.ф.-м.н. Плекану Р.М.; доц., к.б.н.
Осипенку А.П.; асп. Мартишичкіну В.О. та ін. за участь у дискусії
та обговоренні результатів дослідження на тематичних наукових
семінарах відділення ``фізики ядра і елементарних частинок''
кафедри теоретичної фізики.

\end{document}